\documentclass[11pt]{article}
\pdfoutput=1
\usepackage{jheppub}

\newcommand\fft[2]{{\frac{#1}{#2}}}
\newcommand\ft[2]{{\textstyle\frac{#1}{#2}}}
\newcommand\nn{\nonumber}
\newcommand{\qb}{\bar{q}}

\newcommand{\Lzb}{\bar{L}_0}

\newcommand{\hb}{\bar{h}}

\usepackage{booktabs}
\usepackage{slashed}

\usepackage[final]{feynmp}
\makeatletter
\def\endfmffile{%
    \fmfcmd{\p@rcent\space the end.^^J%
        end.^^J%
        endinput;}%
    \if@fmfio
    \immediate\closeout\@outfmf
    \fi
    \ifnum\pdfshellescape=\@ne
    \immediate\write18{mpost \thefmffile}%
    \fi}
\makeatother

\begin{document}

\preprint{}

\title{\boldmath Quantum Corrections to Central Charges and Supersymmetric Casimir Energy in AdS$_3$/CFT$_2$ }

\author[a]{Arash Arabi Ardehali,}
\affiliation[a]{Department of Physics and Astronomy, Uppsala University,\\
Box 516, SE-751 20 Uppsala, Sweden}
\emailAdd{ardehali@physics.uu.se}
\author[b]{Finn Larsen,}
\author[b]{James T. Liu,}
\affiliation[b]{Leinweber Center for Theoretical Physics, Randall Laboratory of Physics,\\
The University of Michigan, Ann Arbor, MI 48109--1040, USA}
\emailAdd{larsenf@umich.edu} \emailAdd{jimliu@umich.edu}
\author[c,d]{and Phillip Szepietowski}
\affiliation[c]{Institute for Theoretical Physics, University of Amsterdam, \\
Science Park 904, 1098 XH Amsterdam, The Netherlands}
\affiliation[d]{Institute for Theoretical Physics and
Center for Extreme Matter and Emergent Phenomena,\\
Utrecht University, Princetonplein 5, 3584 CC Utrecht, the
Netherlands} \emailAdd{p.g.szepietowski@uu.nl}

\abstract{We study the Casimir energy of bulk fields in AdS$_3$ and its relation to subleading terms in the central charge of the dual CFT$_2$.
Computing both sides of the standard CFT$_2$ relation $E=-c/12$ independently we show that
this relation is not necessarily satisfied at the level of
individual bulk supergravity states, but in theories with sufficient
supersymmetry it is restored at the level of bulk supermultiplets.
Assuming only $(0,2)$ supersymmetry (or more), we improve the situation by
relating quantum corrections to the central charge and the
supersymmetric Casimir energy which in turn is related to an index.
These relations adapt recent progress on the AdS$_5$/CFT$_4$
correspondence to AdS$_3$/CFT$_2$ holography. We test our formula
successfully in several examples, including the $(0,4)$ MSW theory
describing classes of 4D black holes and the large $(4,4)$ theory
that is interesting for higher spin holography. We also make
predictions for the subleading central charges in several recently proposed
$(2,2)$ dualities where the CFT$_2$ is not yet well-understood.}

\maketitle \flushbottom


\section{Introduction}

More than 20 years after the advent of the AdS/CFT correspondence,
``tests" of holographic duality rarely challenge the conjecture
itself, but nevertheless regularly motivate investigations in quantum
field theory that are interesting in their own right. Reproducing the trace anomaly presents an important
``test" of holography in this sense. On the boundary, the anomalous
breaking of conformal symmetry due to an inert background geometry
is parametrized by central charges that are usually well understood
even for finite rank $N$ of the gauge group. In the dual bulk theory
the leading contribution to the anomaly for large rank $N$ of the
gauge symmetry is due to a universal boundary term that is classical
in nature \cite{Henningson:1998gx}. For our purposes, the
interesting bulk effect is the term that is at the subleading order
in $1/N$ and due to one-loop contributions from bulk excitations.
The computation of such quantum effects is well understood in simple
cases but the elaborate spectrum required for precision holography
presents challenges. There has been significant progress on this
problem in the case of AdS$_5$/CFT$_4$ correspondence and it has
largely been solved
\cite{Bilal:1999ph,Mansfield:2000zw,Mansfield:2002pa,Ardehali:2013gra,Ardehali:2013xya,Ardehali:2014,Beccaria:2014xda,Ardehali:2015a}.
The purpose of this paper is to address the situation in the case of
AdS$_3$/CFT$_2$ correspondence, focusing on bulk issues.

As a matter of principle, one-loop quantum effects in the bulk
theory include all modes in the theory, including modes from string
theory (or some other UV completion) that might not be
well-understood. Therefore, it is important to identify the settings
where there is sufficient SUSY that it can be justified to ignore
unknown parts of the spectrum, leaving only the known SUGRA states.
Our goal is to achieve agreements in {\it all} situations where the
legitimacy of focusing on SUGRA states is guaranteed by SUSY, though
{\it not necessarily} in situations where SUSY is too weak, except
perhaps for occasional ``accidents" (that might well be due to some
interesting symmetry beyond SUSY that we have not recognized).

This goal may seem modest but we identify some obstacles. A key
challenge is that there is an infinite tower of Kaluza-Klein states
so, effectively, the theory is higher dimensional. Renormalization
of the theory in AdS$_3$ and the subsequent sum over the KK tower
does not necessarily agree with renormalization of the theory in
higher dimensions.

A specific computational scheme was introduced by
Beccaria, Macorini, and Tseytlin (BMT) \cite{Beccaria:2014}. It
amounts to computing the subleading Casimir energy $\delta E$ by
summing over the ground state energy $\fft12\omega$ for full KK
towers of chiral primaries in supergravity with renormalization in
the $\zeta$-function scheme, and then applying the standard CFT$_2$
relation $\delta E=-\delta c/12$ to extract the subleading central
charge. BMT tested their prescription successfully in the cases of
non-chiral theories with $(4,4)$ SUSY.

We apply the BMT prescription to compute $\delta E$ in situations with less
SUSY. We also introduce an independent algorithm to compute $\delta c$
holographically so we can test the relation $\delta E=-\delta c/12$
at the level of individual bulk representations. These tests are largely successful,
but not always so.

For example, in theories with fewer than four supercharges the relation is not
satisfied. However, this ``failure" is not unexpected because, as we show,
in this case long multiplets contribute to both $\delta E$ and $\delta c$.
Unknown massive fields would be organized in such multiplets and so our
supergravity computation is insufficient for an agreement. The AdS/CFT correspondence
does not require agreement for individual bulk fields so, in principle, it could be that the sum over all KK fields would restore the relation
$\delta E=-\delta c/12$. However, even if such agreement could be established it would anyway be satisfying only after also explaining why long multiplets cancel between themselves. It is an additional and independent concern that the $\zeta$-function regularization may not respect supersymmetry.

To address the situation in a principled manner we consider a different approach, inspired
by progress on the AdS$_5$/CFT$_4$ correspondence over the last few
years. Assuming at least $(0,2)$ SUSY, one can define a right-handed
superconformal index $\mathcal{I}_{\mathrm{s.p.}}^R(q)$ which counts
all the bulk single-particle states that are annihilated by one of
the supercharges and hence vanishes when evaluated on long
representations.\footnote{The superconformal index is essentially
the spectral flow of the elliptic genus to the NS-NS sector; see
Section~\ref{sec:Index} for the explicit definitions.}  The main
result of the present paper is a simple relation between the
high-temperature behavior of this single-particle index and the
subleading central charge of the boundary CFT. More precisely, we
show that the one-loop correction to the left-moving bulk central charge is given by
\begin{equation}
\delta
c_L=12\lim_{\beta\to0}\frac{\mathrm{d}}{\mathrm{d}\beta}\mathcal{I}_{\mathrm{s.p.}}^R(q=e^{-\beta}).\label{eq:mainResult}
\end{equation}
The expression subject to the limit $\beta\to 0$ is a meromorphic function of $\beta$ that should be renormalized by omission of pole terms prior to the limit. An analogous expression for $\delta c_R$ applies to theories with $(2,0)$ SUSY.

To arrive at the formula (\ref{eq:mainResult}), we leverage the
notion of supersymmetric Casimir energy $E^{L}_{SUSY}$, related to
$c_L$ via $E^{L}_{SUSY}=-c_L/24$. That a supersymmetric version of
the Casimir energy can be defined from the superconformal index was
first proposed in \cite{Kim:2012ava,Kim:2013nva} at the level of
single-letter indices in free field theory. We apply this idea to
the bulk single-particle index in order to extract the subleading
$E^L_{SUSY},$ and hence the central charge $\delta c_L$.

We successfully test the simple relation (\ref{eq:mainResult}) in Section~\ref{sec:Examples}, in the context of
several well-known AdS$_3$/CFT$_2$ dualities. We also make predictions for the
subleading central charges in several other cases where the bulk index is known but the
boundary CFT is not.

This paper is organized as follows. In Section~\ref{sec:4dEsusy} we
review some aspects of the supersymmetric Casimir energy $E_{SUSY}$
that were previously developed in the context of the AdS$_5$/CFT$_4$
correspondence. It allows a concise expression of the bulk quantum
corrections to the Weyl anomaly coefficients $a$ and $c$ obtained in
\cite{Ardehali:2014,Ardehali:2015a}. This will serve as an
introduction to the techniques we employ in AdS$_3$/CFT$_2$
holography and motivate our formula (\ref{eq:mainResult}).
In Section~\ref{sec:2dcandE} we discuss possible relations between
the bulk Casimir energy of supergravity states in AdS$_3$ and the
central charge of the dual 2D CFT. We refine the discussion by
defining left-moving and right-moving Casimir energies and attempt
to relate these to the corresponding left and right-moving central
charges, emphasizing the relevance of a supersymmetric spectrum to
the success or failure of such relations at the level of individual
bulk multiplets.
In Section~\ref{sec:Index} we derive the supersymmetric Casimir
energy from the superconformal index and compute from it the
left-moving central charge of $\mathcal N =(0,2)$ theories. In order
to provide a consistency check of our approach in this minimal
supersymmetric context we additionally find the subleading {\it
right-moving} central charge $\delta c_R$ by computing one-loop
Chern-Simons levels $\delta k_R$ for the $U(1)_R$ gauge field, which
are related by $\delta c_R = -3 \delta k_R.$ We also discuss
analogous relations in theories with more supercharges. In
Section~\ref{sec:Examples} we test the results from
Section~\ref{sec:Index} in particular examples of AdS$_3$/CFT$_2$
duality, including the $(0,4)$ MSW theory describing classes of 4D
black holes and the large $(4,4)$ theory that is interesting for
higher spin holography. We also offer predictions for quantum
corrections to the central charge in several $(2,2)$ cases where the
supersymmetric index of bulk supergravity states is known but the
dual CFT$_2$ is not. We conclude with a summary and a discussion of
outlooks in Section~\ref{sec:disc}.

\section{Supersymmetric Casimir Energy and the Index in \texorpdfstring{AdS$_5$/CFT$_4$}{AdS5/CFT4}}\label{sec:4dEsusy}

Before turning to AdS$_3$/CFT$_2$ and the justification of the relation (\ref{eq:mainResult}) it is instructive to review the AdS$_5$/CFT$_4$ case.

The supersymmetric partition function $Z_{SUSY}$, defined as the
path integral with appropriate boundary conditions, is proportional to
the index $\mathcal I$, defined in terms of a trace over the states
that respect supersymmetry. The factor of proportionality defines the
supersymmetric Casimir energy $E_{\rm SUSY}$ through \cite{Kim:2012ava,Kim:2013nva,Assel:2014paa,Ardehali:2015hya,Assel:2015nca,Bobev:2015kza}.

\begin{equation}\label{eq:ZIrel}
Z_{\rm SUSY} = e^{-\beta E_{\rm SUSY}} \mathcal I~.
\end{equation}
In view of its definition, $E_{\rm SUSY}$ plays well with supersymmetry, while the conventional
Casimir energy might not.

Like the Casimir energy in 2D, the supersymmetric Casimir energy in 4D is related to the central
charges of the theory. Specifically, it can be written in terms of the Weyl anomaly coefficients $a$ and $c$ of an $\mathcal
N=1$ SCFT as \cite{Assel:2014paa,Ardehali:2015hya,Assel:2015nca,Bobev:2015kza}
\begin{equation}\label{eq:4dEsusyRef}
E_{\rm SUSY} = \frac{8}{3}(b_1+b_2)(a-c) + \frac{8}{27}\frac{(b_1+b_2)^3}{b_1b_2}(3c-2a).
\end{equation}
Here $b_1$ and $b_2$ parametrize families of supersymmetric boundary
conditions or, equivalently, the fugacities $t=e^{(b_1+b_2)\beta}$
and $y=e^{(b_1-b_2)\beta}$ that appear in the superconformal index
introduced in \eqref{eq:SI} below.

The leading semi-classical saddle point determines central charges of $\mathcal O(N^2)$. From the bulk point of view the subleading $\mathcal O(1)$  contributions are quantum corrections. The superconformal index provides a useful regulator for determining these quantum corrections.
In particular, the subleading central
charges---given below in \eqref{eq:acshort}---can be computed by
extracting the subleading supersymmetric Casimir energy from the
single-trace superconformal index \cite{Ardehali:2014,Ardehali:2015a}.

In AdS$_5,$ bulk excitations can be organized into multiplets of the
superconformal algebra. Representations are labeled by the conformal
dimension $\Delta$, a $U(1)_R$ charge $r$, and two $SU(2)$ spins
$j_1$ and $j_2$. The central charges are only quantum corrected by
loops of shortened multiplets
\cite{Ardehali:2013gra,Beccaria:2014xda}. The contribution from a
single short multiplet is given by
\cite{Ardehali:2013xya,Beccaria:2014xda}
\begin{align}\label{eq:acshort}
c_{\rm chiral} = -c_{\rm SLII} &= (-1)^{2j_1+2j_2}\frac{9}{128}(2j_1+1)(\Delta-\ft{r}{2}-1)[(\Delta-\ft{r}2)(\Delta-\ft{r}2-2)+\ft49j_1(j_1+1)+\ft49], \nn\\
a_{\rm chiral} = -a_{\rm SLII} &= (-1)^{2j_1+2j_2}\frac{9}{128}(2j_1+1)(\Delta-\ft{r}{2}-1)[(\Delta-\ft{r}2)(\Delta-\ft{r}2-2)-\ft43j_1(j_1+1)+\ft23],
\end{align}
where chiral multiplets satisfy the shortening condition
\begin{equation}
\Delta = \frac{3r}{2}, \qquad j_2 = 0,
\end{equation}
and semi-long II (SLII) multiplets satisfy
\begin{equation}
\Delta = \frac{3r}{2} +2j_2 + 2.
\end{equation}
Multiplets with conserved currents are included as the special case of SLII multiplets with the additional condition $\frac{3}{2}r - j_1 - j_2 = 0.$ The fact that the contributions are equal and opposite for the two
types of shortened multiplets is consistent with the recombination
rules for forming a long multiplet out of a chiral and an SLII
multiplet along with the fact that long multiplets do not contribute
to the central charges. The central charges also receive  contributions from anti-chiral and SLI multiplets; these are given by \eqref{eq:acshort} with the replacements $j_1 \leftrightarrow j_2$ and $r\rightarrow -r.$

The right-handed $\mathcal N=1$ superconformal index is given by the trace
\begin{equation}\label{eq:SI}
\mathcal I^R(t,y) = \text{Tr} (-1)^F t^{-(\Delta-\fft{r}2)}y^{2j_1}.
\end{equation}
This only receives contributions from states which satisfy $\Delta-\fft32r-2j_2=0$.  As
these states are not present in long multiplets, the index only
receives contributions from shortened representations of the
superconformal algebra.
Henceforth we will drop the $R$ superscript. Analogous results for the left-handed index follow from the replacements $j_1 \leftrightarrow j_2$ and $r\rightarrow - r.$

The theories we consider are effectively free at large $N,$ so
multiparticle states are generated from single particle states by
simple combinatorics. In AdS$_5$/CFT$_4$ this means we can consider
single-particle states in AdS which are dual to single-trace
operators in the CFT. In this case the full multi-particle index
$\mathcal I(t,y)$ is simply related to the single-particle index by
means of the plethystic exponential
\begin{equation}\label{eq:PlethExp}
\mathcal I(t,y) = \exp \left(\sum_{k=1}^\infty\frac{1}{k} \mathcal I_{s.p.}(t^k,y^k)\right),
\end{equation}
where the single-particle index is given by the same trace in \eqref{eq:SI}, except restricted to single-particle states
\begin{equation}
\mathcal I_{s.p.}(t,y) = \text{Tr}_{s.p.}(-1)^F t^{-(\Delta-\frac{r}{2})}y^{2j_1}.
\end{equation}
The single-particle superconformal index for individual short multiplets is
\begin{equation}\label{eq:short4dIndex}
\mathcal I^{\rm chiral}_{s.p.}(t,y) = -\mathcal I^{\rm SLII}_{s.p.}(t,y) = (-1)^{2(j_1+j_2)} \frac{t^{-(\Delta-\frac{r}{2})}\chi_{j_1}(y)}{(1-t^{-1} y)(1-t^{-1} y^{-1})}.
\end{equation}
where
$$
\chi_{j}(y) = \fft{y^{j+\fft12} - y^{-j-\fft12}}{y^{\fft12} - y^{-\fft12}},
$$
is the $SU(2)$ character.

The free-field description allows us to compute the supersymmetric Casimir energy by extracting the term that is linear in $\beta$ from the single-particle index.
In particular, by inserting the index for a shortened multiplet given in \eqref{eq:short4dIndex} into
\begin{equation}\label{eq:4dEsusyfromI}
E_{\rm SUSY} =  -\frac{1}{2}\lim_{\beta\rightarrow 0}\partial_\beta \mathcal I_{s.p.}(t=e^{(b_1+b_2)\beta},y=e^{(b_1-b_2)\beta})\Big|^{\textrm{\footnotesize finite}}.
\end{equation}
we recover the result in \eqref{eq:4dEsusyRef} with $a$ and $c$
given by \eqref{eq:acshort}.\footnote{For the precise numerical
agreement note that the Casimir energy in \eqref{eq:4dEsusyRef} is
written in terms of the full central charge, which receives
contributions also from anti-chiral and SLI multiplets.  These
satisfy $a_{\rm chiral}(\Delta,j,0,r) = a_{\rm
anti-chiral}(\Delta,0,j,-r)$ and $a_{\rm SLII}(\Delta,j_1,j_2,r) =
a_{\rm SLI}(\Delta,j_2,j_1,-r)$ so they introduce an overall factor
of two.} Thus we can utilize the superconformal index as a method of
computing the central charges. This is the main idea that we will
adapt to the context of AdS$_3$/CFT$_2$ in Section~\ref{sec:Index}.

The result in \eqref{eq:4dEsusyfromI} can alternatively be understood as a regularized sum of supersymmetric energies of
states \cite{Kim:2012ava,Kim:2013nva}, with the derivative with
respect to $\beta$ in \eqref{eq:4dEsusyfromI} bringing down the
supersymmetric energy of a single-particle state. In
Appendix~\ref{App:zeta} we discuss the relation between this type of regularization and the
zeta-function regularization in the  context of AdS$_3$/CFT$_2$.
This explains why the term in the single-particle index that is linear in $\beta$ is equivalent to
the $\mathcal O(1)$ contribution to the supersymmetric Casimir energy as defined in \eqref{eq:ZIrel} for large-$N$ theories.

\section{Central Charge and the Casimir Energy in \texorpdfstring{AdS$_3$/CFT$_2$}{AdS3/CFT2}}\label{sec:2dcandE}


In two-dimensional conformal field theories, the standard ({\it
i.e.} non-supersymmetric) Casimir energy is simply related to the
central charge by \cite{Bloete:1986qm,Affleck:1986bv}
\begin{equation}
E_c = - \frac{c}{12}.
\label{eq:Ecc}
\end{equation}
In this section, we explore the connection between these two quantities from a holographic point of view.  As in the previous section, our focus is on the $\mathcal O(1)$ contributions arising from one-loop effects in the bulk.  As reviewed below, both quantities can be obtained from the one-loop partition function, which can be decomposed as a sum over the Kaluza-Klein spectrum.  From this point of view, it was noted in \cite{Beccaria:2014} that the Casimir energy/central charge relation does not generally hold for the holographic contribution of individual states in the bulk.  Nevertheless, since (\ref{eq:Ecc}) must hold for all two-dimensional conformal field theories, it must somehow be recovered after summing over the Kaluza-Klein tower.

Some of the best developed examples of AdS$_3$/CFT$_2$ duality have $\mathcal N=(4,4)$ supersymmetry and \cite{Beccaria:2014} found
that, with this much supersymmetry, the relation (\ref{eq:Ecc}) does hold, after all. It is satisfied on an individual multiplet by multiplet basis, even before the sum over the Kaluza-Klein states is taken. However, AdS$_3$/CFT$_2$ correspondence applies more generally and it is interesting to find criteria for the viability of a holographic description. Thus we explore theories with various amounts of supersymmetry, starting with $\mathcal N=(0,2)$.

\subsection{The Holographic Weyl Anomaly}\label{subsec:bulkCandVeff}

The holographic Weyl anomaly encoded in the leading central charge $c^{(0)}$ can be obtained
from the logarithmically divergent part of the bulk action \cite{Henningson:1998gx}.
For our purposes it is convenient to express the result as
\cite{Brown:1986nw,Kraus:2005}
\begin{equation}
c^{(0)}=6\pi\ell^3\frac{I_{\mathrm{on-shell}}}{\mathrm{vol}(AdS_3)},
\end{equation}
where $I_{\mathrm{on-shell}}$ is the classical action
\begin{equation}
I=\frac{1}{16\pi G_3}\int \sqrt{g}\mathcal{L}_3,
\end{equation}
evaluated on-shell. For Einstein gravity with a cosmological
constant the on-shell Lagrangian is given by $\mathcal
L_3=-R+2\Lambda=4/\ell^2$, which yield the Brown-Henneaux central
charge $c^{(0)}=3\ell/2G_3$.

We can generalize the classical bulk result to include quantum corrections by modifying the action $I\to I+\delta I_{\mathrm{eff}}$ so that the central charge receives a correction
\begin{equation}
\delta c=6\pi\ell^3\frac{\delta
I_{\mathrm{eff}}}{\mathrm{vol}(AdS_3)}.\label{eq:deltaCfromIeff}
\end{equation}
Here we take the one-loop effective action
\begin{equation}
\delta I_{\mathrm{eff}}=(-1)^{2s}\frac{1}{2}\log\det D,
\label{eq:IeffLogDet}
\end{equation}
where $s$ is the spin and $D$ is the appropriate differential operator appearing in the bulk field's kinetic term. The statistical factor $(-1)^{2s}$ takes into account the fact that fermionic and bosonic determinants go in the numerator and denominator of the partition function, respectively.

The one-loop effective action is divergent, and can be regulated by the spectral zeta function.  For an AdS$_3$ representation labeled by $SO(2,2)$
quantum numbers $(\Delta,s)$ we find
\begin{align}
\delta I_{\mathrm{eff}}&=-\frac{1}{2}(-1)^{2s}\zeta'(0,D)\nn\\
&=\frac{1}{6}(-1)^{2s}\nu(\nu^2-3s^2)\log R,\label{eq:effAcZeta}
\end{align}
where $\nu=\Delta-1$, and $R$ is a cut-off scale regularizing the AdS$_3$ volume. In going to the second line we have used the result of \cite{Giombi:2013} (their Eq.~(3.8)) for the spectral zeta function. Combining (\ref{eq:deltaCfromIeff}) with (\ref{eq:effAcZeta}), and using the regulated $\mathrm{vol}(AdS_3)=-2\ell^3\pi\log R$, we arrive at \cite{Beccaria:2014}%
\footnote{Note that our $\delta c$ is denoted in \cite{Beccaria:2014} by $c_{\mathrm{AdS}_3}^+$.}%
\begin{equation}
\delta c(\Delta,s)=-\frac{1}{2}(-1)^{2s}\nu(\nu^2-3s^2)\qquad(\nu=\Delta-1).
\end{equation}
We may also translate the AdS$_3$ labels $(\Delta,s)$ to the
equivalent CFT weights $h$ and $\bar h$, so that
\begin{equation}
\delta c(h,\bar h)=-\frac{1}{2}(-1)^{2s}\nu(\nu^2-3s^2)\qquad(\nu=h+\bar h-1,\quad s=h-\bar h).
\label{eq:deltaCforBulkField}
\end{equation}
Finally, for \emph{massless}
bulk fields which have $\bar{h}$ (resp.\ $h$) equal to zero, a
further ghost contribution should be included so that the combined
contribution to $\delta c_{\mathrm{massless}}$ is $\delta
c(h,0)-\delta c(h,1)$ (resp.\ $\delta c(0,\bar{h})-\delta
c(1,\bar{h})$) \cite{Metsaev:1995re}.

\subsection{The Casimir Energy}

Turning now to the Casimir energy, we take a 2D CFT point of view and start with the partition function
\begin{equation}\label{eq:PF}
Z(q,\qb) = \textrm{Tr}\, (-1)^{F_L+F_R}q^{L_0} \qb^{\Lzb},
\end{equation}
where $L_0$ and $\Lzb$ are left and right Virasoro generators and
the total fermion number can be replaced with the spin of the
representation as $(-1)^{F_L + F_R} = (-1)^{2s} =
(-1)^{2(L_0-\Lzb)}$.  For free fields we can then formally define
\begin{equation}
E_c = \frac{1}{2}\textrm{Tr}(-1)^F(L_0+\bar L_0)=\left.\frac{1}{2} q \frac{d}{dq} Z(q,\qb = q)\right|_{q\to1}.
\label{eq:dqZ}
\end{equation}
The idea expressed by the formula is very basic, a generalization of the ground state energy $\fft12 \hbar\omega$ for the harmonic oscillator. However, as always in quantum field theory, we must address divergences. We find it convenient to introduce a regulator through the substitution $q = e^{-\epsilon}$ which renders \eqref{eq:dqZ} a meromorphic function of $\epsilon$. We subsequently take the limit $q\to 1$ by retaining only the constant term in the Laurent expansion around $\epsilon = 0$. As a concise shorthand for this procedure we write:
\begin{equation}\label{eq:EcZ}
E_c = - \frac{1}{2} \lim_{\epsilon\rightarrow0} \frac{d}{d\epsilon} Z(e^{-\epsilon},e^{-\epsilon})\Big|^{\textrm{\footnotesize finite}}.
\end{equation}
This prescription for extracting the Casimir energy is equivalent to that defined via the spectral zeta function. Indeed, following \cite{Beccaria:2014}, one defines
\begin{equation}
E_c = \frac{1}{2}(-1)^{2s} \sum_n (h_n+\bar h_n) = \frac{1}{2}(-1)^{2s}\zeta(-1),
\end{equation}
where $h_n$ and $\bar h_n$ are eigenvalues of $L_0$ and $\Lzb$ and
\begin{equation}
\zeta(z) = \sum_n \frac{1}{(h_n+\bar h_n)^z} = \frac{1}{\Gamma(z)}\int_{0}^\infty d\beta \beta^{z-1}Z(e^{-\beta},e^{-\beta}).
\end{equation}

One can refine \eqref{eq:dqZ} by taking independent $q$ and $\bar q$ derivatives of the partition function.  In this way, we may define the ``left'' and ``right'' contributions to the Casimir energy
\begin{eqnarray}\label{eq:EcLZ}
E^L_c &=& - \frac{1}{2} \lim_{\epsilon\rightarrow0} \left( \frac{d}{d\epsilon} Z(e^{-\epsilon},\qb) \right)_{\qb = e^{-\epsilon}} \Big|^{\textrm{\footnotesize finite}},\\
\label{eq:EcRZ}
E^R_c &=& - \frac{1}{2} \lim_{\epsilon\rightarrow0} \left( \frac{d}{d\epsilon} Z(q,e^{-\epsilon}) \right)_{q = e^{-\epsilon}} \Big|^{\textrm{\footnotesize finite}}.
\end{eqnarray}
In these two formulae the large round brackets serve to stress that the finite substitutions $\qb\rightarrow e^{-\epsilon}$ and $q\rightarrow e^{-\epsilon}$ are performed after differentiation with respect to $\epsilon$ but before extracting the constant part of the expressions from the
Laurent expansion in $\epsilon$ around $\epsilon=0.$ The prescriptions for $E^L_c$ and $E^R_c$ in \eqref{eq:EcLZ} and \eqref{eq:EcRZ} can also be extracted from the Casimir energy computed from a suitably refined zeta function. However, they do not themselves correspond precisely to the Casimir energy and are instead given by particular derivatives of the refined Casimir energy. Therefore, while it is tempting to relate $E^L_c$ and $E^R_c$ to the left and right-moving central charges, it is not clear that such an identification is justified. However, it follows straightforwardly from the chain rule that the sum
\begin{equation}
E_c = E^L_c + E^R_c,
\end{equation}
reproduces the physical Casimir energy.

Until now, we have not yet specified the one-particle partition function $Z(q,\bar q)$.  For a bulk field dual to an operator with weights $h$ and $\bar h$, it takes the simple form
\begin{equation}\label{eq:Zhbarh}
Z_{h,\hb}(q,\qb) = \frac{q^h\qb^{\hb}}{(1-q)(1-\qb)},
\end{equation}
where the denominator arises from the SL($2,\mathbb{R}$)$\times$ SL($2,\mathbb{R}$) descendents of the highest weight state.  Performing the manipulations above, we find
\begin{align}
E_{c}^L(h,\hb) &= -\frac{1}{48}(-1)^{2(h-\hb)}\left(1-10h+18h^2-8h^3+12h\hb-12h^2\hb-6\hb^2+4\hb^3\right), \nn\\
E_{c}^R(h,\hb) &= -\frac{1}{48}(-1)^{2(h-\hb)}\left(1-10\hb+18\hb^2-8\hb^3+12h\hb-12h\hb^2-6h^2+4h^3\right).
\label{eq:ELR}
\end{align}
Adding these together and replacing $h=\frac{1}{2}(\nu+s+1)$ and $\hb = \frac{1}{2}(\nu+1-s)$ then gives \cite{Beccaria:2014}
\begin{equation}
E_{c}(h,\bar{h})=\frac{1}{24}(-1)^{2s}\nu(2\nu^2-1)\qquad(\nu=h+\bar h-1,\quad s=h-\bar h).
\label{eq:Ec}
\end{equation}
As noted in \cite{Beccaria:2014}, the central charge expression
(\ref{eq:deltaCforBulkField}) and the Casimir energy (\ref{eq:Ec})
in general do not obey the two-dimensional CFT relation $E_c=-c/12$.
This is perhaps an unusual aspect of AdS$_3$/CFT$_2$ holography, and
we will explore this connection in more detail below.  To do so, we
find it convenient to define the would-be central charge
\begin{equation}
\delta\tilde c \equiv -12 E_c=-12(E_c^L+E_c^R),
\end{equation}
obtained from the Casimir energy.  Below, we examine the relation
between these two notions of the central charge for theories with
varying amounts of supersymmetry.

\subsection{Holography for Various Amounts of Supersymmetry}

The finite-dimensional subgroup $SO(2,2)$ of the two-dimensional conformal group splits into left and right components, $SU(1,1)\times SU(1,1)$.  This can be extended by including varying amounts of supersymmetry on the left and the right, independently.  We now turn to a few important cases.

\subsubsection{\texorpdfstring{$\mathcal N=(0,2)$}{N = (0,2)}}

The $\mathcal N=2$ superconformal algebra corresponds to $SU(1,1|1)$, and extends the conformal algebra formed by $L_0$, $L_{\pm1}$ with the supercurrents $G_{\pm1/2}^{\pm}$ and a $U(1)$ current $J_0$.  Lowest weight representations are labeled by the weight $h$ and charge $r$, and unitary representations exist for $h\ge|r|$.  Such representations fall into two categories, namely long for $h>|r|$ and short for $h=|r|$.  The latter may be classified as either chiral for $h=r$ or anti-chiral for $h=-r$. The content of these representations are given by
\begin{align}
\mbox{long:}&\qquad |h,r\rangle\oplus|h+\ft12,r+\ft12\rangle\oplus|h+\ft12,r-\ft12\rangle\oplus|h+1,r\rangle,\nn\\
\mbox{chiral:}&\qquad|h,h\rangle\oplus|h+\ft12,h-\ft12\rangle,\nn\\
\mbox{antichiral:}&\qquad|h,-h\rangle\oplus|h+\ft12,\ft12-h\rangle.
\end{align}

For the $\mathcal N=(0,2)$ case, we take a single irreducible representation $|h\rangle$ on the left and tensor it with $|\bar h,\bar r\rangle$ on the right.  The resulting representations are then classified by the right-moving superalgebra, and the result for the Casimir energies and central charge are shown in Table~\ref{tbl:(0,2)}.  As shown in the table, the would-be central charge $\delta\tilde c$ computed from the Casimir energy does not agree with the holographic calculation of $\delta c$.

\begin{table}[t]
\begin{center}
\begin{tabular}{c|ll}
$\mathcal N=(0,2)$&long ($\bar h>|\bar r|$)&short ($\bar h=|\bar r|$)\\
\hline
$E_c^L$&$-\ft1{8}\bar h$&$-\ft1{48}(1-6h+6h^2+3\bar h-6\bar h^2)$\\
$E_c^R$&$-\ft1{16}(1-2h-4\bar h)$&$-\ft1{32}(1-8\bar h+8\bar h^2-2h+8h\bar h)$\\
$\delta\tilde c$&$\ft34(1-2h-2\bar h)$&$\ft18(5+12h^2+12\bar h^2-18h-18\bar h+24h\bar h)$\\
\hline
$\delta c$&$\ft34(1-2h+2\bar h)$&$\ft18(5+12h^2-12\bar h^2-18h-6\bar h+24h\bar h)$
\end{tabular}
\caption{The left and right Casimir energies, $E_c^L$ and $E_c^R$, would-be central charge $\delta\tilde c$ and central charge $\delta c$ for $\mathcal N=(0,2)$ multiplets. Note that $\delta\tilde c\ne\delta c$ for both short and long multiplets. All entries should be multiplied by the spin-dependent factor $(-1)^{2(h-\bar h)}$.  }
\label{tbl:(0,2)}
\end{center}
\end{table}

\subsubsection{\texorpdfstring{$\mathcal N=(2,2)$}{N = (2,2)}}

Although $\mathcal N=(0,2)$ supersymmetry is insufficient to give
agreement between $\delta\tilde c$ and $\delta c$, it turns out that
a match is obtained for $\mathcal N=(2,2)$ superconformal theories.
Here, since both sides are supersymmetric, we may tensor together
either long or short multiplets on both sides.  This gives rise to
the four possibilities shown in Table~\ref{tbl:(2,2)}.  Importantly,
the long-long representations do not contribute to either the
Casimir energy nor the central charge.  Moreover, $E_c^L$ only
receives contributions from shortened representations on the right,
while $E_c^R$ only receives contributions from shortened
representations on the left.  The fact that short representations
suffice for certain computations underpins our ability to employ
two-dimensional superconformal indices for efficiently packaging
those computations below.

\begin{table}[t]
\begin{center}
\begin{tabular}{c|llll}
$\mathcal N=(2,2)$&long-long&long-short&short-long&short-short\\
\hline
$E_c^L$&$0$&$-\ft1{16}$&$0$&$-\ft1{32}(1-4h)$\\
$E_c^R$&$0$&$0$&$-\ft1{16}$&$-\ft1{32}(1-4\bar h)$\\
$\delta\tilde c$&$0$&$\ft34$&$\ft34$&$\ft34(1-2h-2\bar h)$\\
\hline
$\delta c$&$0$&$\ft34$&$\ft34$&$\ft34(1-2h-2\bar h)$
\end{tabular}
\caption{The left and right Casimir energies, $E_c^L$ and $E_c^R$,
would-be central charge $\delta\tilde c$ and central charge $\delta
c$ for $\mathcal N=(2,2)$ multiplets.  All multiplets have
$\delta\tilde c=\delta c$. All entries should be multiplied by the
spin-dependent factor $(-1)^{2(h-\bar h)}$.  } \label{tbl:(2,2)}
\end{center}
\end{table}

\subsubsection{Small \texorpdfstring{$\mathcal{N}=(0,4)$}{N = (0,4)}}\label{subsubsec:N=(0,4)}

We now turn to theories with $\mathcal N=4$ supersymmetry. Here we
have to make a distinction between the ``small" and the ``large" $\mathcal N=4$.
The small algebra contains $PSU(1,1|2)$ as its finite dimensional
subalgebra and the large $\mathcal A_\gamma$ contains $D(2,1|\alpha)$ .

We start with the small $\mathcal N=4$ algebra, with
the finite dimensional subalgebra generated by $L_0$, $L_{\pm1}$,
$SU(2)$ currents $J_0^i$ and supercurrents $G_{\pm1/2}^a$, $\tilde
G_{\pm1/2}^a$ transforming as doublets under $SU(2)$.  In addition,
the central charge is related to the level $k$ of the affine $SU(2)$
via $c=6k$.

Lowest weight representations of the small $\mathcal N=4$ algebra
are built from a state $|h,j\rangle$, where $j$ now labels the
$SU(2)$ representation.  Once again, unitary representations exist
for $h\ge j$, with saturation of the inequality corresponding to
shortened representations. Complete representations for the small
$\mathcal N=(0,4)$ case are then obtained by tensoring a
representation $|h\rangle$ on the left with an $\mathcal N=4$
representation $|\bar h,\bar j\rangle$ on the right. The resulting
representations are short for $\bar h=\bar j$ and long for $\bar
h>\bar j$.

The resulting Casimir energies and central charges are given in Table~\ref{tbl:(0,4)}.  We find that the small $\mathcal N=(0,4)$ supersymmetry is sufficient for the would-be central charge derived from the Casimir energy $\delta\tilde c$ to coincide with the holographic central charge $\delta c$.

\begin{table}[t]
\begin{center}
\begin{tabular}{c|ll}
\multicolumn{1}{c}{small}\\
$\mathcal N=(0,4)$&long ($\bar h>\bar j$)&short ($\bar h=\bar j$)\\
\hline
$E_c^L$&$0$&$-\ft1{24}(1-6h+6h^2)$\\
$E_c^R$&$0$&$\ft1{8}\bar h(1-2h)$\\
$\delta\tilde c$&$0$&$\ft12(1+6h^2+6h\bar h-6h-3\bar h)$\\
\hline
$\delta c$&$0$&$\ft12(1+6h^2+6h\bar h-6h-3\bar h)$
\end{tabular}
\caption{The left and right Casimir energies, $E_c^L$ and $E_c^R$, would-be central charge $\delta\tilde c$ and central charge $\delta c$ for small $\mathcal N=(0,4)$ multiplets labeled by $|h;\bar h,\bar j\rangle$.  All multiplets have $\delta\tilde c=\delta c$. All entries should be multiplied by the spin-dependent factor $(-1)^{2(h-\bar h)}$.   }
\label{tbl:(0,4)}
\end{center}
\end{table}

\subsubsection{Small \texorpdfstring{$\mathcal N=(4,4)$}{N = (4,4)}}

Turning to the small $\mathcal N=(4,4)$ case, we can build representations by tensoring together either long or short $\mathcal N=4$ representations on the left and on the right.  Since the Casimir energy and central charge vanishes identically whenever there is an $\mathcal N=4$ long representation, whether on the left or the right, only the short-short case will contribute non-trivially.  The results are summarized in Table~\ref{tbl:(4,4)}.  The short-short result can also be obtained by decomposing the multiplet into  $\mathcal N=(2,2)$ representations and using the results shown in Table~\ref{tbl:(2,2)}.  Once again, we find that there is sufficient supersymmetry that $\delta\tilde c=\delta c$. This is the result previously reported in \cite{Beccaria:2014}.

\begin{table}[t]
\begin{center}
\begin{tabular}{c|llll}
\multicolumn{1}{c}{small}\\
$\mathcal N=(4,4)$&long-long&long-short&short-long&short-short\\
\hline
$E_c^L$&$0$&$0$&$0$&$\ft14h$\\
$E_c^R$&$0$&$0$&$0$&$\ft14\bar h$\\
$\delta\tilde c$&$0$&$0$&$0$&$-3(h+\bar h)$\\
\hline
$\delta c$&$0$&$0$&$0$&$-3(h+\bar h)$
\end{tabular}
\caption{The left and right Casimir energies, $E_c^L$ and $E_c^R$, would-be central charge $\delta\tilde c$ and central charge $\delta c$ for small $\mathcal N=(4,4)$ multiplets.  All multiplets have $\delta\tilde c=\delta c$.  A ll entries should be multiplied by the spin-dependent factor $(-1)^{2(h-\bar h)}$.  }
\label{tbl:(4,4)}
\end{center}
\end{table}

\subsubsection{Large \texorpdfstring{$\mathcal N=(0,4)$}{N = (0,4)}}

We now consider the large $\mathcal N=4$ cases.  The
large $\mathcal N=4$ superalgebra contains, in addition to Virasoro,
two affine $SU(2)$'s and a $U(1)$ algebra.  The fermionic generators
include dimension-3/2 supercurrents and dimension-1/2 fields, both
transforming as $(2,1)+(1,2)$ under the two $SU(2)$'s.  The two
affine $SU(2)$'s have levels $k^+$ and $k^-$, and the central charge
is given by $c=6k^+k^-/(k^++k^-)$.  It is customary
to define the parameters
\begin{equation}
\gamma=\fft{k^-}{k^++k^-},\qquad\alpha=\fft\gamma{1-\gamma}=\fft{k^-}{k^+}.
\end{equation}
For a holographic point of view, we are mostly interested in the finite subalgebra $D(2,1|\alpha)$ with bosonic component $SL(2;\mathbb R)\times SU(2)\times SU(2)$. It admits unitary representations whenever~\cite{deBoer:1999}
\begin{equation}
h\ge\gamma j^++(1-\gamma)j^-,
\end{equation}
Once again, representations split into long and short, with the latter saturating this bound.  Using (\ref{eq:ELR}) and (\ref{eq:deltaCforBulkField}), we then arrive at the results shown in Table~\ref{tbl:large(0,4)}. We see that generally the large $\mathcal N=(0,4)$ is insufficient
to ensure $\delta\tilde c=\delta c$ even though it has four supersymmetries. However, the equality applies for some representations,
notably those with $\bar j^+=\bar j^-$.

\begin{table}[t]
\begin{center}
\begin{tabular}{c|ll}
\multicolumn{1}{c}{large}\\
$\mathcal N=(0,4)$&long&short\\
\hline
$E_c^L$&$0$&$-\ft1{24}[1-6h+6h^2+6\gamma(1-\gamma)(\bar j^+-\bar j^-)^2]$\\
$E_c^R$&$0$&$-\ft1{8}(\bar j^+-\bar j^-)[(1-2\gamma)(1-2h)-4\gamma(1-\gamma)(\bar j^+-\bar j^-)]$\\
$\delta\tilde c$&$0$&$\ft12[1+6h^2-6h +3(\bar j^+-\bar j^-)((1-2\gamma)(1-2h)-2\gamma(1-\gamma)(\bar j^+-\bar j^-))]$\\
\hline
$\delta c$&$0$&$\ft12[1+6h^2-6h +3(\bar j^+-\bar j^-)((1-2\gamma)(1-2h)+2\gamma(1-\gamma)(\bar j^+-\bar j^-))]$
\end{tabular}
\caption{The left and right Casimir energies, $E_c^L$ and $E_c^R$, would-be central charge $\delta\tilde c$ and central charge $\delta c$ for large $\mathcal N=(0,4)$ multiplets labeled by $|h;\bar h,\bar j^+,\bar j^-\rangle$.  We find $\delta\tilde c\ne\delta c$ except for special cases such as the representations $\bar j^+=\bar j^-$ or the limits $\gamma\to0,1$. All entries should be multiplied by the spin-dependent factor $(-1)^{2(h-\bar h)}$.  }
\label{tbl:large(0,4)}
\end{center}
\end{table}

\subsubsection{Large \texorpdfstring{$\mathcal{N}=(4,4)$}{N = (4,4)}}\label{subsec:largeDeltaC}

Finally, we examine the large $\mathcal N=(4,4)$ case, or more precisely representations of $D(2,1|\alpha)\times D(2,1|\bar\alpha)$.  Again, we tensor together either long or short representations on the left and the right.  The results summarized in Table~\ref{tbl:large(4,4)} show that
in this case $\delta\tilde c=\delta c$.

Although we allow for different parameters on the left and the right in the discussion here, the case we have in mind is string theory on AdS$_3\times S^3\times S^3\times S^1$ which has $\alpha=\bar\alpha$, equivalent to $\gamma=\bar\gamma$. Furthermore, the BPS spectrum of Type II supergravity on AdS$_3\times S^3\times S^3\times S^1$ only contains states with $j^+=j^-$ and $\bar j^+=\bar j^-$ \cite{Eberhardt:2017a,Baggio:2017kza}.  In this special case all states in the Kaluza-Klein reduction therefore give vanishing contributions to $E_c^L$, $E_c^R$ and $\delta c$.

\begin{table}[t]
\begin{center}
\begin{tabular}{c|llll}
\multicolumn{1}{c}{large}\\
$\mathcal N=(4,4)$&long-long&long-short&short-long&short-short\\
\hline
$E_c^L$&$0$&$0$&$0$&$-\ft14(1-2\gamma)(j^+-j^-)$\\
$E_c^R$&$0$&$0$&$0$&$-\ft14(1-2\bar\gamma)(\bar j^+-\bar j^-)$\\
$\delta\tilde c$&$0$&$0$&$0$&$3[(1-2\gamma)(j^+-j^-)+(1-2\bar\gamma)(\bar j^+-\bar j^-)]\kern-4pt$\\
\hline
$\delta c$&$0$&$0$&$0$&$3[(1-2\gamma)(j^+-j^-)+(1-2\bar\gamma)(\bar j^+-\bar j^-)]\kern-4pt$
\end{tabular}
\caption{The left and right Casimir energies, $E_c^L$ and $E_c^R$, would-be central charge $\delta\tilde c$ and central charge $\delta c$ for large $\mathcal N=(4,4)$ multiplets. All multiplets have $\delta\tilde c=\delta c$. All entries should be multiplied by the spin-dependent factor $(-1)^{2(h-\bar h)}$.  }
\label{tbl:large(4,4)}
\end{center}
\end{table}

\subsection{How Much Supersymmetry is Sufficient?}

As we have seen, the holographic dual to the CFT relation $\delta E_c = -\delta
c/12$ (equivalent to $\delta\tilde c=\delta c$ in the notation used in this section) is not generally upheld
on a state by state basis in the bulk.  In particular, it fails for bulk states with no supersymmetry at all and it similarly fails for
$\mathcal N=(0,2)$ multiplets.

On the other hand, it holds for all theories with four or more supercharges, with the exception of large
$\mathcal N=(0,4)$ supersymmetry where restriction to representations with $\bar j^+=\bar j^-$ is needed. The restriction needed in the latter case may or may not be significant; it appears to be satisfied for
all known models with large $\mathcal N=4$ supersymmetry in the
bulk. This indicates that the standard Casimir energy may not be a useful
tool for computing subleading central charges in AdS$_3$/CFT$_2$
duals with too little supersymmetry.

Of course, even in theories with fewer supercharges, we still expect  $\delta E_c = -\delta c/12$ to be valid once we sum
over the complete bulk spectrum. However, the mechanism for this equality is far from obvious in the cases where it does not follow from supersymmetry applied to individual multiplets.

\section{Supersymmetric Casimir Energy in \texorpdfstring{AdS$_3$}{AdS3}}\label{sec:Index}

In this section we propose an improved method for computing quantum corrections to the bulk central charge in AdS$_3$/CFT$_2$
holography. The linchpin is a supersymmetric version of the Casimir energy due to bulk fluctuations. This supersymmetric Casimir energy
arises naturally in the context of partition functions that preserve some amount of supersymmetry which are closely related to the superconformal index.

As we will see, in the context of chiral $\mathcal N = (0,2)$ or $\mathcal N = (0,4)$ theories this approach will provide a useful
tool in computing the left-moving subleading central charge $\delta c_L$. In theories where both chiralities preserve two or more supercharges we are
able to compute quantum corrections to both the left and right-moving central charges in a manifestly supersymmetric fashion.

\subsection{The Superconformal Index and the Supersymmetric Casimir Energy}
In 2D CFTs with  $\mathcal N = (0,2)$ supersymmetry the
superconformal index is defined as
\begin{equation}
\mathcal I_R (q) = \text{Tr} (-1)^{2(L_0-\bar L_0)} q^{L_0} \bar q^{\bar L_0 - \bar J_0},
\label{eqn:genindex}
\end{equation}
where the trace is over all states in the theory on the cylinder
$S^1\times \mathbb R.$ The index only receives contributions from
states satisfying $\bar L_0 = \bar J_0$. These only appear in
shortened (chiral) multiplets with respect to a particular
supercharge in the superconformal algebra. The index is thus
independent of $\bar q$ and depends only on the BPS spectrum of the
theory. We use a subscript $R$ to emphasize that the index localizes
on multiplets that are shortened with respect to one of the right
moving supercharges.

One can similarly define an anti-chiral index that localizes on
states that are shortened with respect to the other supercharge.
Also, in theories with a left-moving supersymmetry one can
straightforwardly define analogous left-handed indices that localize
onto shortened states of the left-moving algebra.

The index also has an interpretation as the partition function given by the Euclidean path-integral for the theory on $S^1\times S^1,$ with fermions satisfying periodic boundary conditions around the temporal circle. The trace and the partition function
representations of the index are related by an overall factor
\begin{equation}\label{eq:ZEIR}
Z_R(q) = e^{-\beta E^L_{SUSY}}\mathcal I_R(q).
\end{equation}
Here $q= e^{-\beta}$, where $\beta$ is defined by the ratio of the
length of the Euclidean time circle to the length of the spatial
circle. We will often refer to $\beta$ as an inverse temperature
even though we have given fermions periodic boundary conditions. The
quantity $E^L_{SUSY}$ in the exponent of the prefactor is the
supersymmetric Casimir energy. We use a superscript $L$ in order to
emphasize that in free field theory it is essentially a sum of
eigenvalues of the left-moving $L_0$ over all the states killed by
the right-moving supercharge ({\it c.f.}
Eq.~(\ref{eq:sumExpressionEL}) below). Assuming a unique ground
state and an appropriate gap above it so that the index satisfies
$\mathcal I_R(q=0) = 1$, then $E^L_{SUSY}$ can be extracted from the
$\beta \rightarrow \infty$ limit of the partition function.

The supersymmetric Casimir energy has a universal structure determined by the anomaly polynomial of the SCFT. In particular, the Casimir energy determined from the right-handed partition function through \eqref{eq:ZEIR} is given simply by the left-moving central charge \cite{Bobev:2015kza}
\begin{equation}
E_{\rm SUSY}^L = -\frac{c_L}{24}.
\end{equation}

In theories that admit a holographic dual at large $c$ the central charge decomposes
into a sum of a leading contribution derived from the classical supergravity action and a
subleading contribution that is due to one-loop determinants of fields in the bulk. Similarly, the supersymmetric Casimir energy can
be decomposed as
\begin{equation}
E_{SUSY}^L = E^L_{(0)} + E^L_{(1)}.
\end{equation}
The first term $E^L_{(0)}$ is a classical contribution from the bulk
on-shell action evaluated on the $AdS_3$ background with an
appropriate supersymmetric regularization prescription
\cite{Genolini:2016sxe,Genolini:2016ecx}.\footnote{While references
\cite{Genolini:2016sxe,Genolini:2016ecx} only consider the cases
of AdS$_4$ and AdS$_5,$ we expect that similar results also
hold for AdS$_3.$} The second term $E^L_{(1)}$ arises from one-loop
contributions to the supersymmetric bulk partition function. In
notation that is a natural generalization of the previous section
one then has the relations
\begin{equation}
E^{L}_{(0)} = -\frac{c^{(0)}_L}{24}
\end{equation}
and
\begin{equation}\label{eq:EL1}
E^L_{(1)} = - \frac{\delta c_L}{24}.
\end{equation}
%
Our focus is on
the computation of $E_{(1)}^L$ and will use \eqref{eq:EL1} to relate
it to $\delta c_L.$

In order to extract $\delta c_L$ we will utilize a particular property of the index at large $N$. Rewriting \eqref{eq:ZEIR} as
\begin{equation}
\mathcal I_R(q) = e^{\beta E^L_{SUSY}}Z_R(q),
\end{equation}
and evaluating $Z_R(q)$ holographically for large $c$ limit we
see that, since the large-$c$ index does not scale with $c,$ the
leading behavior at large $c$ arising from $E^L_{(0)}$ should cancel
against the corresponding contribution to the partition function at large $c$. We can then simplify this
to\footnote{Our estimates all refer to the leading saddle point at
large $c$ and quantum fluctuations around it. Additional contributions from subleading saddles can be
neglected here because their suppression is of order $e^{-c}$.}
\begin{equation}\label{eq:Zone-loopI}
\mathcal I^{\text{large-}c}_R(q) = e^{\beta E^L_{(1)}}Z^\text{one-loop}_R(q),
\end{equation}
where $\mathcal I^{\text{large-}c}_R(q)$ is the $c\rightarrow \infty$ limit of the index and $Z^\text{one-loop}_R(q)$ refers to the contribution to the partition function from one-loop determinants of all bulk fields. At one-loop, the contributions to $Z_R(q)$ correspond to free fields in the bulk which have an interpretation as generalized free fields in the CFT.

In a high-temperature (small-$\beta$) expansion, the logarithm of
the partition function for free fields has no term linear in $\beta$
\cite{Plunien:1986ca,Plunien:1987fr,Gibbons:2006ij}. Because of its
generalized free field interpretation, we expect that $\ln
Z_R^{\text{one-loop}}$ similarly has no term linear in $\beta$ in
its small-$\beta$ expansion. We can therefore compute the quantum
correction $E^L_{(1)}$ to the supersymmetric Casimir energy from the
index at large $c$ and then infer the corresponding quantum
correction $\delta c_L$ from \eqref{eq:EL1}. We substantiate this
procedure in Appendix \ref{App:zeta}, by demonstrating that the
linear-in-$\beta$ term of $\ln \mathcal I_R^{\text{large}-c}$ can be
reduced to a sum over left-moving energies of free single-particle
states (annihilated by the right-handed supercharge), and so can
naturally be identified with $E^L_{(1)}$ in the bulk.

\subsection{The Single-particle Index}\label{subsec:cL02}
In theories with a holographic description the index at large $c$ is naturally expressed in terms of contributions from single-trace operators in the theory.
For theories with $\mathcal N= (0,2)$ we express it as
\begin{equation}\label{eq:IstR}
\mathcal I^{s.t.}_{R}(q) = \text{Tr}_{s.t.}(-1)^{2(L_0-\bar L_0)} q^{L_0}\bar q^{\bar L_0 - \bar J_0}.
\end{equation}
Each single-trace operator in the CFT is dual to a supergravity field, so in the bulk the trace can be thought of as being over single-particle states and we can replace ``s.t." above with ``s.p." While we will not use it in the following, we note for completeness that the full index \eqref{eqn:genindex} at large $c$ is constructed from the single-particle result by taking the plethystic exponential
\begin{equation}\label{eq:IRPE}
\mathcal I_R^{\text{large-}c}(q) = \exp\left(\sum\limits_{k=1}^\infty\frac{1}{k}\mathcal I^{s.p.}_{R}(q^k)\right),
\end{equation}
which takes the multi-particle contributions into account .

The supersymmetric Casimir energy can be computed from the index in
the same way that the standard Casimir energy was extracted from the
partition function in \eqref{eq:EcZ}, {\rm viz.}
\begin{equation}\label{eq:EL1index}
E^L_{(1)} = -\frac{1}{2} \lim_{\beta\rightarrow 0}\frac{d}{d\beta}
\mathcal I^{s.p.}_R(q=e^{-\beta})\Big|^{\textrm{\footnotesize
finite}}.
\end{equation}
We show in Appendix \ref{App:zeta} that this relation is equivalent to computing
\begin{equation}
E^L_{(1)} = \sum(-1)^{2s}\  \frac{1}{2}h_n\label{eq:sumExpressionEL}
\end{equation}
using zeta-function regularization. It therefore has a natural
interpretation as a Casimir energy (of the supersymmetric states) in
the bulk, as expected.

Combining the supersymmetric Casimir energy \eqref{eq:EL1index} with
(\ref{eq:EL1}) we have our main result
\begin{equation}\label{eq:mainResultSec3}
\delta c_L = 12 \lim_{\beta\rightarrow 0}\frac{d}{d\beta}
\mathcal I^{s.p.}_R(q=e^{-\beta})\Big|^{\textrm{\footnotesize
finite}}.
\end{equation}
This provides an algorithm for computing the quantum corrections to the left-moving
central charge in $\mathcal N= (0,2)$ supersymmetric theories with
a holographic dual.

To find explicit formulae, we organize bulk states into multiplets of the $\mathcal N = (0,2)$ algebra, which come in two types: short and long. Short chiral multiplets satisfy $\bar L_0 = \bar J_0$ and contribute to the index as
\begin{equation}\label{eq:I(0,2)}
\mathcal I^{s.p.}_{R,h,\bar h}(q) = (-1)^{2(h-\bar h)}\frac{q^h}{1-q},
\end{equation}
where $h$ and $\bar h$ are the left and right-moving weights of the operator and the denominator arises from descendant contributions. Importantly, short anti-chiral multiplets (with $\bar L_0 = - \bar J_0$) and long multiplets (with $\bar L_0 > |\bar J_0|$) give vanishing contributions to \eqref{eq:IstR}.

Inserting the single-particle index \eqref{eq:I(0,2)} for a chiral $\mathcal N=(0,2)$
multiplet into (\ref{eq:EL1index}), we find
\begin{equation}\label{eq:chiral_Esusy}
E^L_{(1)}(h,\bar h) = -\frac{1}{24}(-1)^{2(h-\bar h)}(1-6h+6h^2),
\end{equation}
which gives
\begin{equation}\label{eq:cL}
\delta c_{L}(h,\bar h) =(-1)^{2(h-\bar h)}( 1 - 6h+6h^2).
\end{equation}
While for $\mathcal N= (0,2)$ theories we do not
yet have a direct way of corroborating this result, when we apply it
and its corresponding right-moving counterpart in theories with both
left and right-moving supercharges we can compare with the Weyl
anomaly results for $\delta c$. As we will see, the result in
\eqref{eq:cL} corresponds to the sum of contributions to $\delta
c_L$ from a chiral and the corresponding CP conjugate anti-chiral
multiplet.

\subsection{Quantum Corrections from One-loop Chern-Simons Levels}

In an effort to support the veracity of our formula
\eqref{eq:cL} for the central charge $\delta c_L$ in theories with only $\mathcal N = (0,2)$
supersymmetry it is useful to compute the right-moving central charge $\delta c_R.$ This necessarily involves another method.

The central charge is related to the level $k_R$ of the right-moving $U(1)_R$ current by
\begin{equation}\label{eq:cRkR}
\delta c_R = -3 \delta k_R,
\end{equation}
where $\delta k_R$ is the bulk one-loop contribution to coefficient of the Chern-Simons term
\begin{equation}\label{eq:CSLag}
\delta \mathcal L \propto \frac{1}{4\pi }\delta k_R\,  A_R\wedge dA_R.
\end{equation}
Holographic anomaly matching gives the $F_R\wedge F_R$ term in the $U(1)_R$ anomaly. 

The quantum correction $\delta k_R$ can be computed from one-loop vacuum polarization
diagrams in three-dimensional flat space, extracting the contribution to the Chern-Simons term in the bulk. The topological nature of the Chern-Simons term implies that the
flat space result is valid also in curved space. Explicit computations presented in
Appendix \ref{App:CS} give\footnote{We derive these results for spins $s=\ft12,1$ and $\ft32$. While we do not consider spins greater than $\ft32$, the simple extrapolation used here gives results consistent with expectations. It would be interesting
to verify it by an explicit calculation. See also \cite{Grimm:2018weo} for an
independent computation of these Chern-Simons levels using an index
theorem. }
\begin{equation}\label{eq:keff}
\delta k_R(s,\bar r) = -(-1)^{2s}\, 2s \,\bar r^2,
\end{equation}
for a field of spin $s=h-\bar h$ and right-moving $U(1)_R$-charge $\bar r$. For an analogous calculation of one-loop Chern-Simons terms in five dimensions, see
\cite{Bonetti:2013ela}.

Summing the result \eqref{eq:keff} over the states in an $\mathcal N = (0,2)$ multiplet we find
\begin{align}\label{eq:kRshort}
\delta k^\text{short}_R(h,\bar h)  &= -\fft14(-1)^{2(h-\bar h)}(1 - 4\bar h^2 - 2h - 2\bar h + 8 h \bar h ), \\
\label{eq:kRlong}
\delta k^\text{long}_R(h,\bar h)  &=  -\fft1{2}(-1)^{2(h-\bar h)}(1  - 2h + 2\bar h).
\end{align}
and then \eqref{eq:cRkR} gives the corresponding values for $\delta c_R$. Combining these results with our formula for $\delta c_L$,
taking the degeneracies arising from chiral and anti-chiral multiplets into
account, we arrive at values for the total central charge
\begin{equation}
\delta c = \frac{1}{2}(\delta c_L + \delta c_R),
\end{equation}
for both short and long multiplets. The values for the Weyl anomaly
$\delta c$ obtained this way agree with the results in
Table~\ref{tbl:(0,2)}, summed over a chiral multiplet and its CP
conjugate anti-chiral multiplet. Importantly, we see also that
$(0,2)$ supersymmetry is enough to ensure that long supermultiplets
in the bulk do not contribute to $\delta c_L$. These observations
provide evidence supporting our formula \eqref{eq:EL1index}
computing $E^L_{(1)}$ from the right-handed index.

However, note
that there is no {\it a priori} obvious way to regularize a sum of
Chern-Simons levels over a Kaluza-Klein tower (without recourse to the higher-dimensional embedding as in \cite{Grimm:2018weo}) as in the case of
quantities computed directly from the index, such as $\delta c_L$.
Therefore, utilizing the results in \eqref{eq:kRshort} and
\eqref{eq:kRlong} to compute such a KK sum should be done with care.

\subsection{Higher Amounts of Supersymmetry}
In the remainder of this section we present results for cases with higher amounts of supersymmetry. We will see that the case of $\mathcal N = (0,4)$ is similar to $\mathcal N = (0,2)$ discussed above and that the story simplifies when there is both a left-moving and a right-moving supersymmetry.

\subsubsection{Small \texorpdfstring{$\mathcal N = (0,4)$}{N = (0,4)}}\label{subsec:cL04}

In the case of $\mathcal N = (0,4)$ supersymmetry, the index of a
short multiplet is given by the same expression \eqref{eq:I(0,2)} as
in the $\mathcal N=(0,2)$ case. This means that the result for
$\delta c_L$ in \eqref{eq:cL} is also valid for short multiplets
with $\mathcal N = (0,4)$ supersymmetry. Indeed, it matches exactly
with the result from $E_c^L$ in Table~\ref{tbl:(0,4)}, upon
identifying $E_c^L$ in Section~\ref{sec:2dcandE} with the
$E^L_{(1)}$ defined in this section. This indicates that, for
theories with at least four ``small'' supercharges, the somewhat
{\it ad hoc} prescription pursued in Section~\ref{sec:2dcandE} is
equivalent to the more principled method advanced in the current
section, at least at the level of individual multiplets.

Note that for $\mathcal N=(0,4)$ we do not need to sum the results
from Table~\ref{tbl:(0,4)} over chiral and anti-chiral sectors as we
did in the $\mathcal N=(0,2)$ case. This is because a short
multiplet of $\mathcal N =(0,4)$ contains an entire $SU(2)$
multiplet as its highest weight state. In particular, for each state
in the highest weight representation the corresponding CP conjugate
state is also included and one does not need to add the two together
as in the case with $\mathcal N = (0,2)$.

Like in the $\mathcal N = (0,2)$ case, we can check that our result
for $\delta c_L$ is consistent with the Weyl anomaly $\delta c$
found in Section~\ref{sec:2dcandE} and presented in
Table~\ref{tbl:(0,4)}. For this comparison we compute the
Chern-Simons levels for $(0,4)$ multiplets and find
\begin{align}\label{eq:kR04}
\delta k^\text{short}_R(h,\bar h)  &=(-1)^{2(h-\bar h)}\bar h(1 - 2h), \\
\label{eq:kRlong04} \delta k^\text{long}_R(h,\bar h)  &= 0,
\end{align}
with the corresponding central charges again $\delta c_R = -3\delta k_R$.  Just as $\delta c_L$ found here was consistent with $E^L_c$ in
Table~\ref{tbl:(0,4)}, the result for $\delta k_R^\text{short}$ is consistent with $E^R_c$ in that table.
With $\mathcal N=(0,4)$ supersymmetry the long multiplets offer no further check because,
unlike the $\mathcal N=(0,2)$ case, the contribution from long multiplets to $\delta k_R$ vanishes.

\subsubsection{Large \texorpdfstring{$\mathcal N = (0,4)$}{N = (0,4)}}\label{subsec:cLlarge04}

With large $\mathcal N = (0,4)$ supersymmetry, the index of a short
multiplet is again given by the same expression \eqref{eq:I(0,2)} as
in the $\mathcal N=(0,2)$ case, and hence the result for $\delta
c_L$ in \eqref{eq:cL} is also valid for short multiplets with large
$\mathcal N = (0,4)$ supersymmetry.

Interestingly, this result does not match with the one found using
the ordinary Casimir energy $E_c^L$ and presented in
Table~\ref{tbl:large(0,4)}, except for representations with
$\bar{j}^+=\bar{j}^-$ (or $\gamma=0,1$). Moreover, the discrepancies
occur precisely in the situations where we had already identified
problems with the CFT$_2$ relation $\delta E_c = - \delta c/12$. It
is hard to escape the conclusion that the principled method using
supersymmetric Casimir energy and a relation to an index is correct,
while the prescription in Section~\ref{sec:2dcandE} is unreliable.

In this situation it is particularly important that we can check that the result for $\delta c_L$ is consistent
with the quantum correction $\delta c$ to the holographic Weyl anomaly given in Table~\ref{tbl:large(0,4)}, by performing an independent computation in Chern-Simons theory. The result for the Chern-Simons levels of the $U(1)_R$ reported in \eqref{eq:JofN=2} are
\begin{align}\label{eq:kRlarge04}
\delta k_R^{\text{short}}(h,\bar h,\bar j_+,\bar j_-) &= (-1)^{2(h-\bar h)}(\bar j_- - \bar j_+)((1-2\gamma)(1-2h)-2\gamma(1-\gamma)(\bar j_+-\bar j_-)),\\
\delta k_R^{\text{long}}(h,\bar h,\bar j_+,\bar j_-) &= 0,
\end{align}
and again corresponds to central charge $\delta c_R =
-3\delta k_R$. We find agreement with the central charge $\delta c$ but generally not with the (presumably erroneous) result $\delta \tilde{c}$ that was deduced from the ordinary Casimir energy $E_c^L$.

As in the small $\mathcal N=(0,4)$
case, with large $\mathcal N=(0,4)$ the contribution from long multiplets to $\delta k_R$
vanishes.\footnote{This computation proceeds most
easily if one first computes the sum over the $SU(2)\times SU(2)$ states in a representation of $SL(2,R)\times SU(2)^2,$ giving
\begin{equation}\sum\limits_{(\bar{j}^+,\bar{j}^-)}\bar{r}^2=\frac{1}{3}(1+2\bar{j}^+)(1+2\bar{j}^-)[\gamma^2
\bar{j}^+(1+\bar{j}^+)+(1-\gamma)^2 \bar{j}^-(1+\bar{j}^-)]. \nn\end{equation}}

\subsubsection{\texorpdfstring{$\mathcal N = (2,2)$}{N = (2,2)}}\label{subsec:cL22}

When there is supersymmetry on both the left and right we can utilize the existence of a right-moving index $\mathcal I_R$ as well as a left-moving index $\mathcal I_L$ to compute both $\delta c_L$ and $\delta c_R$. The results for $\delta c_L$ and $\delta c_R$ are presented in
Table~\ref{tbl:(2,2)SUSY}. We find independent agreements with the chiral Casimir energies $E_c^L$ and $E_c^R$ reported in Table~\ref{tbl:(2,2)}, after summing the latter over a chiral multiplet plus its CP conjugate anti-chiral multiplet. This implies agreement also for the holographic Weyl anomaly

\begin{table}[t]
    \begin{center}
        \begin{tabular}{c|llll}
            $\mathcal N=(2,2)$&long-long&long-short&short-long&short-short\\
            \hline
            $\delta c_L$&$0$&$3$&$0$&$\ft3{2}(1-4h)$\\
            $\delta c_R$&$0$&$0$&$3$&$\ft3{2}(1-4\bar h)$\\
            \hline
            $\delta c$&$0$&$\ft3{2}$&$\ft3{2}$&$\ft3{2}(1-2h-2\bar h)$
        \end{tabular}
        \caption{The left and right central charges $\delta c_L$ and $\delta c_R$ as derived from the SUSY Casimir energy and the central charge $\delta c$ for $\mathcal N=(2,2)$ multiplets. The line $\delta c$ agrees with a sum of the results from Table~\ref{tbl:(2,2)} over a chiral plus its CP conjugate anti-chiral multiplet. All entries should be multiplied by the spin-dependent factor $(-1)^{2(h-\bar h)}$.}
        \label{tbl:(2,2)SUSY}
    \end{center}
\end{table}

\subsubsection{Small and Large \texorpdfstring{$\mathcal N = (4,4)$}{N = (4,4)}}\label{subsec:cL44}

With four supercharges on both the left and the right, we only find
non-zero contributions to $\delta c_L$ and $\delta c_R$ from
short-short multiplets. The results for the small ${\cal N}=(4,4)$
are summarized in Table~\ref{tbl:(4,4)SUSY} and those for the large
${\cal N}=(4,4)$ are similarly given in
Table~\ref{tbl:large(4,4)SUSY}. Both are completely consistent with
the results from Section~\ref{sec:2dcandE}, reported in
Table~\ref{tbl:(4,4)} and Table~\ref{tbl:large(4,4)}, respectively.

\begin{table}[t]
    \begin{center}
        \begin{tabular}{c|llll}
        \multicolumn{1}{c}{small}\\
            $\mathcal N=(4,4)$&long-long&long-short&short-long&short-short\\
            \hline
            $\delta c_L$&$0$&$0$&$0$&$-12h$\\
            $\delta c_R$&$0$&$0$&$0$&$-12 \bar h$\\
            \hline
            $\delta c$&$0$&$0$&$0$&$-6(h+\bar h)$
        \end{tabular}
        \caption{The left and right central charges $\delta c_L$ and $\delta c_R$ as derived from the SUSY Casimir energy and the central charge $\delta c$ for $\mathcal N=(4,4)$ multiplets. The line $\delta c$ agrees with the result from Table~\ref{tbl:(4,4)}. All entries should be multiplied by the spin-dependent factor $(-1)^{2(h-\bar h)}$.}
        \label{tbl:(4,4)SUSY}
    \end{center}
\end{table}

\begin{table}[t]
\begin{center}
\begin{tabular}{c|llll}
\multicolumn{1}{c}{large}\\
$\mathcal N=(4,4)$&long-long&long-short&short-long&short-short\\
            \hline
            $\delta c_L$&$0$&$0$&$0$&$6[(1-2\gamma)(j^+-j^-)]$\\
            $\delta c_R$&$0$&$0$&$0$&$6[(1-2\bar\gamma)(\bar j^+-\bar
j^-)]$\\
            \hline
            $\delta c$&$0$&$0$&$0$&$3[(1-2\gamma)(j^+-j^-)+(1-2\bar\gamma)(\bar j^+-\bar
j^-)]\kern-4pt$
\end{tabular}
\caption{The left and right central charges $\delta c_L$ and $\delta
c_R$ as derived from the SUSY Casimir energy and the central charge
$\delta c$ for large $\mathcal N=(4,4)$ multiplets. The line $\delta
c$ corresponds to the result from Table~\ref{tbl:large(4,4)}.  All
entries should be multiplied by the spin-dependent factor
$(-1)^{2(h-\bar h)}$.} \label{tbl:large(4,4)SUSY}
\end{center}
\end{table}

\section{Quantum Corrected Central Charge for Specific Dualities}\label{sec:Examples}

In this section we apply our formula for the quantum correction to
the bulk central charge to full-fledged AdS$_3$/CFT$_2$ dualities in
string theory. In these examples the AdS$_3$ matter content is
qualitatively more complicated than the individual multiplets
considered in the previous sections. We must consider infinite KK
towers, corresponding to supergravity fields in higher dimensions.
The resulting divergences present challenges that are addressed
correctly by our formula, at least in the examples we consider.

We first study cases where the full central charge $c$ is known from
the boundary description; the challenge would then be to reproduce
the $O(1)$ term in the central charge, denoted $\delta c$, from the
bulk description.

In the last two subsections we move on to less-understood dualities
where we will make predictions for the $O(1)$ piece of the central
charges of the yet-to-be-discovered finite-$c$ boundary CFTs.

\subsection{\texorpdfstring{AdS$_3\times S^3\times K3$}{AdS3 x S3 x K3}}

Type IIB theory on AdS$_3\times S^3\times K3$ has small
$\mathcal{N}=(4,4)$ supersymmetry. The dual CFT is a $(K3)^{Q_1
Q_5+1}/S(Q_1 Q_5+1)$ superconformal sigma model \cite{Vafa:1995}.
The seed theory with target space $K3$ has central charge
$c_L=c_R=6$ and the full symmetric orbifold has central charge
$c_L=c_R=6(Q_1 Q_5+1)$. The expansion parameter around the classical
saddle point is
$\frac{G_{\mathrm{AdS}_3}}{\ell_{\mathrm{AdS}_3}}\propto 1/Q_1 Q_5$
\cite{Strominger:1997nh}. The CFT$_2$ therefore has quantum
corrections $\delta c_L=\delta c_R=6$.

The input to the AdS$_3$ computation of $\delta c$ is the
supersymmetric index. The KK towers comprising the supergravity
spectrum was computed by de~Boer \cite{deBoer:1998,deBoer:1998gen}
who presented it in various forms including the one-particle
index\footnote{This is the function $s(1,q,1)$ in eq.~(5.10) of
\cite{deBoer:1998gen} except that, since we are considering the
``one''-particle index, we do not include the $+1$ vacuum
contribution included in the reference.}
\begin{equation}
\mathcal{I}^{\mathrm{s.p.}}_R(q)=\mathcal{I}^{\mathrm{s.p.}}_L(q)=1+\frac{44\sqrt{q}+28q}{1-q}.
\end{equation}
Applying the master formula (\ref{eq:mainResult}) and its
left-moving analogue we arrive at
\begin{equation}
\delta c_L=\delta c_R=-\frac{864}{\beta^2}+6+O(\beta)\quad (\text{as
}\beta\to0).\label{eq:deltaCK3y}
\end{equation}
The divergence in the $\beta\to0$ limit is due to the infinite sum
over towers of Kaluza-Klein modes. Regularizing the divergence by
expanding around $\beta=0$ and keeping the finite piece we obtain
the desired result $\delta c_L=\delta c_R =6$. Our regularization is
manifestly supersymmetric and similar to the one employed
successfully in AdS$_5$/CFT$_4$ \cite{Ardehali:2014,Ardehali:2015a}.
The divergence could also be tamed more traditionally via a
zeta-function regularization \cite{Beccaria:2014}.

\subsection{\texorpdfstring{AdS$_3\times S^3\times T^4$}{AdS3 x S3 x
T4}}\label{subsec:T4}

This case is also a duality with small $\mathcal{N}=(4,4)$ supersymmetry.

The CFT$_2$ description is a $\sigma$-model on the target space
$\tilde{T}^4\times(T^4)^{Q_1 Q_5}/S(Q_1 Q_5)$. The decoupled
$\tilde{T}^4$ in the target space corresponds to the quantum Higgs
branch of the D1-D5 gauge theory \cite{Witten:1997}. From the
viewpoint of the U($Q_1$) gauge theory which flows to the sigma
model in the IR (see {\it e.g.} \cite{Sax:2014}), the $\tilde{T}^4$
represents an $\mathcal{N}=(4,4)$ hypermultiplet decoupled from the
adjoint representation (see {\it e.g.} Section~4 of
\cite{David:2002}). The $\tilde{T}^4$ corresponds in the gravity
picture to two pairs of $(1/2,0)_s+(0,1/2)_s$ singletons living on
the boundary of AdS$_3$. These two singletons are equivalent to the
$\tilde{T}^4$ sigma model on the boundary. Therefore, as far as the
$\tilde{T}^4$ factor is concerned, bulk and boundary theory are
identical and their central charges match trivially:
$c_R=c_L=6$.\footnote{This situation is analogous to the one-loop
Weyl anomaly matching for the U($N$) $\mathcal{N}=4$ SYM and its
bulk dual containing a ``doubleton'' living on the AdS$_5$ boundary.
Once the Weyl anomaly matching of the decoupled U($1$) of the
boundary and the doubleton of the bulk is established (essentially
through their equivalence), it remains to show that the SU($N$)
$\mathcal{N}=4$ SYM and the bulk theory without the doubleton have
identical Weyl anomaly.}

The more interesting part of the CFT$_2$ is the $(T^4)^{Q_1
Q_5}/S(Q_1 Q_5)$ orbifold. The central charges are $c_L=c_R = 6Q_1
Q_5$. Since the expansion parameter again is
$\frac{G_{\mathrm{AdS}_3}}{\ell_{\mathrm{AdS}_3}}\propto 1/Q_1 Q_5$
\cite{Strominger:1997nh} the central charges of the CFT$_2$ are not
corrected by quantum effects $\delta c_L=\delta c_R=0$.

The supersymmetric index for the excitations in the AdS$_3$ is
largely trivial, due to the high symmetry of $T^4$. Once the
$(1/2,0)_s+(0,1/2)_s$ singletons are removed from the bulk spectrum
the indices turn out to be\footnote{Some authors add a $+1$ vacuum
contribution to this index, so it vanishes ({\it c.f.}
\cite{Maldacena:1999tor}). We avoid that in order to consider a
``one''-particle index. The plethystic exponentiation of
(\ref{eq:T4index}) yields a vanishing multi-particle index, as
expected.}
\begin{equation}
\mathcal{I}^{\mathrm{s.p.}}_R(q)=\mathcal{I}^{\mathrm{s.p.}}_L(q)=-1.
\label{eq:T4index}
\end{equation}
The prescription (\ref{eq:mainResult}) thus gives the bulk quantum
correction $\delta c_L=\delta c_R=0$, in agreement with the boundary
dual.

Collecting the two parts of the analysis we find the quantum
corrections $\delta c_L=\delta c_R=6$ for both bulk and boundary
theory, with the nontrivial contributions from singletons and the
overall $\tilde{T}^4$, respectively. BMT achieved this agreement in
\cite{Beccaria:2014} using different methods.

\subsection{\texorpdfstring{AdS$_3\times S^3\times S^3\times S^1$}{AdS3 x S3 x S3 x S1}}

This case is a large $\mathcal{N}=(4,4)$ duality. The dual CFT is
the $Q_1 Q_5^+$-fold symmetric orbifold of a 2D CFT seed theory
referred to as the $\mathcal{S}_\kappa$ theory
\cite{Eberhardt:2017b} (and also \cite{Gukov:2004,Tong:2014} for
earlier works). For comparison with other references note that we
take $\kappa\equiv\frac{Q_5^-}{Q_5^+}-1$. The central charges of
this CFT$_2$ are given by \cite{Eberhardt:2017b}
\begin{equation}
c_L=c_R=\frac{6Q_1 Q_5^- Q_5^+}{Q_5^- +Q_5^+}.
\end{equation}
In the semi-classical gravity regime, the D-brane charges $Q_1$ and
$Q_5^+$, as well as the D5-flux $Q_5^-$, are simultaneously taken to
infinity as $Q_1\propto Q_5^+\propto Q_5^-\to\infty$. Moreover,
there is a constraint that $\frac{Q_5^-}{Q_5^+}\in \mathbb{Z}^{>1}$.
However,  the expansion parameter is
$\frac{G_{\mathrm{AdS}_3}}{\ell_{\mathrm{AdS}_3}}\propto 1/Q_1
Q^+_5$ \cite{Gukov:2004} so the leading quantum correction vanishes
$\delta c_L=\delta c_R=0$.

In order to compute $\delta c_{L,R}$ from the bulk we start from the
BPS spectrum of KK supergravity in this case \cite{Eberhardt:2017a,Baggio:2017kza}
\begin{equation}
\bigoplus_{j\in \frac{1}{2}\mathbb{Z}^{\ge 0}}^{\infty}
([j,j]_s+[j+\frac{1}{2},j+\frac{1}{2}]_s)\otimes
([j,j]_s+[j+\frac{1}{2},j+\frac{1}{2}]_s),\label{eq:specS3S3S1}~.
\end{equation}
The notation $[j^+,j^-]_s$ indicates a short$_{j^+,j^-}$ representation
of $D(2,1|\alpha)$. See the appendix for the details of the
multiplet structure.

It is straightforward to check that each KK level $j$ in the
spectrum (\ref{eq:specS3S3S1}) gives a vanishing contribution to the
supergravity one-particle index. Since the $j=0$ level includes also
the vacuum state whose contribution to the index we should not add,
the total supergravity single-particle index comes out
\begin{equation}
\mathcal{I}^{\mathrm{s.p.}}_R(q)=\mathcal{I}^{\mathrm{s.p.}}_L(q)=-1.
\end{equation}
Then (\ref{eq:mainResult}) establishes the $0=0$ match, very
similarly to the $\mathcal{N}=(4,4)$ duality of the $(T^4)^{Q_1
Q_5}/S(Q_1 Q_5)$ sigma-model.

In this case, instead of going through the index, we could observe
that both on the left and on the right, the short multiplets have
$j^+=j^-$. However, Subsection~\ref{subsec:largeDeltaC} established
that such short$\times$short multiplets do not contribute to $\delta
c_{L,R}$. The $0=0$ agreement between quantum corrections in CFT$_2$
and in AdS$_3$ therefore follows effortlessly.

\subsection{M-theory on \texorpdfstring{AdS$_3\times S^2\times CY_3$}{AdS3 x S2 x CY3}}

These compactifications are interesting because they describe large classes of 4D black holes \cite{MSW:1997}. Moreover,
the duality has $\mathcal{N}=(0,4)$ supersymmetry and is the only chiral example we study in this paper.

The boundary CFT is a sigma model with left-moving central charge \cite{MSW:1997}
\begin{equation}
c_L=P^3+c_2\cdot P-3,
\end{equation}
where $P$ is a very ample divisor in $CY_3$ and $P^3$ is the triple
self-intersection number of the divisor. We have subtracted the $+3$
contribution of the three translational zero-modes of the underlying
M5-brane from the result of \cite{MSW:1997}. This subtraction is
analogous to removing the decoupled $\tilde{T}^4$ from the boundary
target space in the duality of Subsection~\ref{subsec:T4}.

Because the expansion parameter is
$\frac{G_{\mathrm{AdS}_3}}{\ell_{\mathrm{AdS}_3}}\propto 1/P^3$
\cite{Maldacena:1997h}, we have $\delta c_L=-3$. The term that is
linear in $P$ is due to higher-derivative corrections in bulk
\cite{Kraus:2005}, rather then one-loop effects; so it is not our
concern in this work.

Since we have supersymmetry on the right, in the bulk we can compute
the right-handed index of the supergravity theory, and extract
$\delta c_L$ from the index. To do so, we need the single-particle
index of the KK supergravity. It is given in eq.~(6.12) of
\cite{Kraus:2006p}) as
\begin{equation}
\mathcal{I}^{\mathrm{s.p.}}_R(q)= 2(h^{1,1}-h^{1,2})\frac{q}{(1-q)^2}-3\frac{q}{1-q},
\end{equation}
where $h^{i,j}$ are the hodge numbers of the $CY_3$. Applying our master formula
(\ref{eq:mainResult}) we get
\begin{equation}
\delta
c_L=\frac{-48(h^{1,1}-h^{1,2})}{\beta^3}+\frac{36}{\beta^2}-3,
\end{equation}
with the finite piece (as $\beta\to 0$) exactly reproducing the
expected boundary result.

Replacing the $CY_3$ with $T^6$ or $K3\times T^2$ gives more subtle
examples of AdS$_3$/CFT$_2$. We leave the holographic study of
quantum corrections in these examples to the future.

\subsection{\texorpdfstring{AdS$_3\times (S^3\times T^4)/G$}{AdS3 x (S3 x T4)/G} and \texorpdfstring{AdS$_3\times(S^3\times K3)/\mathbb{Z}_2$}{AdS3 x (S3 x
K3)/Z2}}\label{subsec:Eberhardt}

These cases are proposed $\mathcal{N}=(2,2)$ dualities
\cite{Eberhardt:2017c}. The dual CFTs are not yet known at finite
central charge. Therefore instead of presenting matchings, in this
section we will be making predictions.

There are eight dualities in this class: seven dualities involving
orbifolds of $S^3\times T^4$, and one duality involving a
$\mathbb{Z}_2$ orbifold of $S^3\times K3$. In all the eight cases
the orbifold group $G$ rotates the $S^3$ such that a great circle
$S^1$ remains fixed, but the action on the $M_4$ ({\it i.e.} $T^4$
or $K3$) is free so that the internal geometry $(S^3\times M_4)/G$
is smooth. In the first seven cases $T^4/G$ is a Hyperelliptic
Surface (HS), while in the eighth case $K3/\mathbb{Z}_2$ is the
Enriques surface (ES). The proposed infinite-$c$ dual CFTs are
infinite-fold symmetric product orbifolds with seed target spaces HS
or ES, respectively.

For the seven HS cases the bulk one-particle indices are\footnote{
Reference \cite{Eberhardt:2017c} reports zero because it adds a $+1$
contribution from the vacuum.}

\begin{equation}
\mathcal{I}^{\mathrm{s.p.}}_R(q)=\mathcal{I}^{\mathrm{s.p.}}_L(q)=-1,
\end{equation}
just like the case of AdS$_3\times S^3\times T^4$ without the extra
singletons, or the case of AdS$_3\times S^3\times S^3\times S^1$.
Hence our prescription~(\ref{eq:mainResult}) gives the quantum correction
\begin{equation}
\delta c_{L}^{HS}=\delta c_{R}^{HS}=0,
\end{equation}
from the bulk side. Of course, this is disregarding possible bulk singletons that
would be dual to decoupled factors in the boundary sigma model.

For the ES case, the bulk one-particle indices are nontrivial
\begin{equation}
\mathcal{I}^{\mathrm{s.p.}}_R(q)=\mathcal{I}^{\mathrm{s.p.}}_L(q)=\frac{22\sqrt{q}+14q}{1-q}~.
\end{equation}
where again we removed a $+1$ vacuum contribution from the result reported in
\cite{Eberhardt:2017c}. Applying our master formula (\ref{eq:mainResult}) we get
\begin{equation}
\delta c_{L}^{ES}=\delta
c_{R}^{ES}=-\frac{432}{\beta^2}+3,\label{eq:deltaCK3x}
\end{equation}
with the finite part (as $\beta\to 0$) giving our prediction for the
boundary theory: $\delta c_{L}^{ES}=\delta c_{R}^{ES}=3$.

\subsection{\texorpdfstring{AdS$_3\times (S^3\times T^4)/D_n$}{AdS3 x (S3 x T4)/Dn}}

As in the previous subsection, these cases are proposed
$\mathcal{N}=(2,2)$ dualities \cite{Datta:2017}. The dual CFTs are
not yet known at finite central charge, so in this subsection we
will again be making predictions rather than presenting agreements.

There are a total of eight distinct cases. In the notation of \cite{Datta:2017}, there are two for the $D_1$
group (the $D_1^{(1)}$ and $D_1^{(2)}$ cases), and six more for $D_2^{(1)}$, $D_2^{(2)}$, $D_3^{(1)}$, $D_3^{(2)}$,
$D_4$, and $D_6$. The internal
geometry $(S^3\times T^4)/D_n$ is singular in all the cases, so the bulk spectrum is not just KK modes from supergravity in higher dimensions, it also has a part coming from the twisted
sector of IIB string theory. The proposed infinite-$c$ dual CFTs are
infinite-fold symmetric product orbifolds with seed target space
$T^4/D_n$.

For the two dualities with $D_1$ group, the bulk one-particle indices are simply $-1$, just like the
case of AdS$_3\times S^3\times T^4$ without the extra singletons, the case
of AdS$_3\times S^3\times S^3\times S^1$, or the seven HS examples. Therefore,
just as in those cases (\ref{eq:mainResult}) gives
\begin{equation}
\delta c_{L}^{D_1}=\delta c_{R}^{D_1}=0,
\end{equation}
for the quantum correction computed from the bulk side. Of course, this is assuming no extra bulk
singletons that would be dual to possible decoupled factors in the boundary sigma model.

For the remaining six dualities the superconformal index for single particle excitations in the bulk are nontrivial.
It is given by eq.~(6.33) of \cite{Eberhardt:2017b} as
\begin{equation}
\mathcal{I}^{\mathrm{s.p.}}_R(q)=\frac{4(1+a)q^{1/4}+22\sqrt{q}+4(1+a)q^{3/4}+(18+4a)q}{1-q},
\end{equation}
after setting the parameter in that work $z$ to zero, and again removing the $+1$ vacuum
contribution included there. The parameter $a$ takes different values $\{1,-1/2,-1,-1,1/4,-1/2\}$ for
the six distinct bulk geometries that we are considering. Our prescription (\ref{eq:mainResult}) for
the quantum correction to the central charge gives
\begin{equation}
\delta c_{L}=\delta
c_{R}=-\frac{144(4+a)}{\beta^2}+6(1+\frac{a}{2}).\label{eq:cDnOrbs}
\end{equation}
The finite piece, evaluated for the appropriate values of $a$, gives our
predictions
\begin{equation}
\delta c_{L}=\delta c_{R}=6(1+a/2)=\{9, 9/2, 3, 3,
27/4,9/2\}.\label{eq:c2comma2listed}
\end{equation}
for the six boundary theories that are based on orbifold groups $D_2^{(1)}$, $D_2^{(2)}$, $D_3^{(1)}$, $D_3^{(2)}$, $D_4$, $D_6$.

\section{Summary and Outlook}\label{sec:disc}

The main result (\ref{eq:mainResult}) of this article is an
efficient prescription for the holographic computation of quantum
corrections to central charges in the supersymmetric AdS$_3$/CFT$_2$
correspondence. Our one-loop ({\it i.e.}
$\mathcal{O}(G_{\mathrm{AdS}_3}^0)$) formula follows from the
relation between the central charges and the supersymmetric Casimir
energies, as extracted from the high-temperature behavior of the
bulk one-particle superconformal indices.

For theories with non-chiral supersymmetry where the left- and right-moving central charges are equal, our results agree with
those of BMT \cite{Beccaria:2014}, at the level of individual bulk multiplets. However, when considering complete Kaluza-Klein towers
of states our method is advantageous because our regularization of the infinite KK sums is completely unambiguous and manifestly
supersymmetric.

We also leveraged the notion of supersymmetric Casimir energy to
motivate our formula even in cases with chiral supersymmetry. With
``small'' $\mathcal{N}=(0,4)$ SUSY our results again coincide with
those of BMT \cite{Beccaria:2014} at the level of individual bulk
multiplets, but not necessarily for entire KK towers. For multiplets
of $\mathcal N=(0,2)$ or even multiplets of ``large''
$\mathcal{N}=(0,4)$ (for representations with $\bar{j}^+\neq
\bar{j}^-$) we find disagreement.

Although it is possible that agreement would be restored after summing over the whole bulk spectrum, it is not obvious that this can be realized without a supersymmetric prescription for the sum. It would therefore be interesting to test our formula  \eqref{eq:mainResult} for the
KK spectra in instances of $\mathcal{N}=(0,2)$ and $\mathcal{N}=(0,4)$ AdS$_3$/CFT$_2,$ such
as those in \cite{Haghighat:2015ega,Couzens:2017way,Couzens:2017nnr}.

Among the concrete outcomes of the present work were the predictions
in Subsection~\ref{subsec:Eberhardt} for the quantum corrections to
central charges in the dualities recently discovered by Eberhardt
\cite{Eberhardt:2017c}. Initial investigations seem to confirm our
predictions, and reveal novel aspects of AdS$_3$/CFT$_2$ at the
quantum level \cite{EberhardtPC}.

A more ambitious question left for future work is whether a version
of our main result (\ref{eq:mainResult}) can be found that applies
at finite-$N$. Such a formula would relate the exact (left-handed)
central charge of any 2D SCFT to its full (right-handed)
superconformal index. A simple finite-$N$ version would be
\begin{equation}
c_L\overset{?}{=}12\lim_{\beta\to0}\frac{\mathrm{d}}{\mathrm{d}\beta}{\rm PL}\big(\mathcal{I}^R(q=e^{-\beta})\big),\label{eq:generalized}
\end{equation}
where ${\rm PL}$ refers to the plethystic logarithm and
$\mathcal{I}^R$ is the full right-handed SCFT index. A simple
calculation shows that this proposal implies that the term that is
linear in $\beta$ in the asymptotic small-$\beta$ expansion of
$\ln\mathcal{I}^R$ is equal to $-\beta c_L/24$.

Similar finite-$N$ prescriptions were conjectured for 4d SCFTs in
\cite{Ardehali:2015a}, but were later found to be invalid for
SO($3$) SQCD with two flavors \cite{Ardehali:2015c}. This example
shows that such finite-$N$ formulas do not necessarily apply when
the Rains function of the 4d SCFT has flat directions
\cite{Ardehali:2015c}. Such flat directions appear in turn related
to unlifted Coulomb branches in the crossed channel---{\it i.e.} on
$R^3\times S^1$ \cite{Ardehali:2015c}.

Analogously, we expect that (\ref{eq:generalized}) may be violated
when certain gap conditions on the spectrum of the 2D SCFT are not
met. With appropriate gap conditions, on the other hand, one can
establish the proposal (\ref{eq:generalized}). For example, for
$\mathcal{N}=(2,2)$ SCFTs with a gap in the spectrum above $L_0=J_0$
and with finite degeneracies (so that the ``low-temperature''
asymptotics of the index is dominated by the contribution of the
states with $L_0=J_0$), it is a simple exercise to show that the
modular properties of the elliptic genus imply
(\ref{eq:generalized}). It is an interesting prospect that,
ultimately, such gap conditions might have connections with the
sparsity criteria (such as \cite{Hartman:2014oaa,Benjamin:2015hsa})
for a 2D SCFT having a holographic dual in the first place.

\begin{acknowledgments}
    We would like to thank M.~Beccaria, F.~Benini, L.~Eberhardt, H.~het Lam, K.~Mayer, D.~McGady, S.~Murthy,
    O.~Sax, and S.~Vandoren for helpful discussions and correspondences related to this
    project. The work of A.A.A. is supported by the Knut and Alice
    Wallenberg Foundation under grant Dnr KAW 2015.0083. The work of P.S. is part of the Delta ITP consortium, a program of the NWO that
    is funded by the Dutch Ministry of Education, Culture and Science (OCW). This work was
    supported in part also by the U.S. Department of Energy under grant
    DE-SC0007859.
\end{acknowledgments}

\appendix

\section{Some 2D Superconformal Representation Theory}

In this appendix, we present a brief overview of superconformal representations
used in the body of the paper.  Our starting point is the $\mathcal N=2$ superconformal algebra, which can be further extended to yield the small and large $\mathcal N=4$ algebras.  The two-dimensional conformal algebra decomposes into left- and right-moving sectors, and for the most part we will focus on a single copy.  However, we  also report the contribution of individual representations to the right-handed superconformal index
\begin{equation}
\mathcal{I}_R(q)=\mathrm{Tr}(-1)^{2L_0-2\bar{L}_0}q^{L_0}
\bar{q}^{\bar{L}_0-\bar{J}_0}.
\label{eq:rhindex}
\end{equation}
When doing so, we will consider theories with equal amounts of supersymmetry on both sides.  Note that this index assumes at least $\mathcal N=(0,2)$ supersymmetry, with $L_0$ and $J_0$ defined by the $\mathcal N=2$ algebra given below.  The contributions to
$\tilde{\mathcal{I}}_R(q)=\mathrm{Tr}(-1)^{2L_0-2\bar{L}_0}q^{L_0}
\bar{q}^{\bar{L}_0+\bar{J}_0}$ and the left-handed indices (assuming supersymmetry on the left) are
obtained similarly.

\subsection{The \texorpdfstring{$\mathcal{N}=2$}{N=2} Superconformal Algebra}

The $\mathcal{N}=2$ superconformal algebra corresponds to SU($1,1|1$), and it has a bosonic
subgroup SU$(1,1)\times$U($1$). The global part of the algebra is
given by $L_0$, $L_{\pm1}$, $G^+_{\pm1/2}$, $G^-_{\pm1/2}$, $J_0$,
with corresponding (anti-)commutation relations
\begin{align}
[L_0,L_{\pm1}]&=\mp L_{\pm1},&[L_1,L_{-1}]&=2L_0,\nn\\
{}[L_0,G^+_{\pm1/2}]&=\mp\ft12G_{\pm1/2}^+,&
[L_0,G^-_{\pm1/2}]&=\mp\ft12G_{\pm1/2}^-,\nn\\
{}[L_1,G_{-1/2}^\pm]&=G_{1/2}^\pm,&
[L_{-1},G_{1/2}^\pm]&=-G_{-1/2}^\pm,\nn\\
{}\{G_{\pm1/2}^+,G_{\pm1/2}^-\}&=2L_{\pm1},&
\{G_{\pm1/2}^+,G_{\mp1/2}^-\}&=2(L_0\pm J_0),\nn\\
{}[J_0,G_{1/2}^\pm]&=\pm\ft12G_{1/2}^\pm,&
[J_0,G_{-1/2}^\pm]&=\pm\ft12G_{-1/2}^\pm.
\end{align}
Lowest weight representations are built from the state $|h,r\rangle$
where $h$ and $r$ are the $L_0$ and $J_0$ eigenvalues, respectively.
The full representation is then obtained by acting with a combination of the
creation operators $L_{-1}$, $G_{-1/2}^+$ and $G_{-1/2}^-$.

\subsubsection*{Representations of the $\mathcal N=2$ Superconformal Algebra}

Unitary irreducible representations exist for $h\ge|r|$.  The case
$h>|r|$ gives rise to an $\mathcal N=2$ long representation built
from four $\mathcal N=0$ representations
\begin{equation}\label{eq:N=02long}
|h,r\rangle,\ |h+\ft12,r+\ft12\rangle=G_{-1/2}^+|h,r\rangle,\
|h+\ft12,r-\ft12\rangle=G_{-1/2}^-|h,r\rangle,\
|h+1,r\rangle=G_{-1/2}^+G_{-1/2}^-|h,r\rangle.
\end{equation}
When $h=|r|$, we end up with a shortened representation.  The chiral representation corresponds to starting
with $|h,h\rangle$, in which case $G_{-1/2}^+$ creates a null state.
As a result, the chiral multiplet is generated by $L_{-1}$ and
$G_{-1/2}^-$.  The $\mathcal N=0$ decomposition is
\begin{equation}\label{eq:N=02ch}
|h,h\rangle,\qquad|h+\ft12,h-\ft12\rangle=G_{-1/2}^-|h,h\rangle.
\end{equation}
Finally, the antichiral multiplet is built from
\begin{equation}
|h,-h\rangle,\qquad|h+\ft12,\ft12-h\rangle=G_{-1/2}^+|h,-h\rangle.
\end{equation}

\subsubsection*{Contribution to the Index}

Focusing on the right-handed index (\ref{eq:rhindex}), it is easy to see that long and antichiral representations on the right do not contribute.  Hence only chiral multiplets on the right contribute, giving a factor $(-1)^{-2\bar{h}}$ for the $|\bar h,\bar h\rangle$ multiplet.  As a result, we have
\begin{equation}
\mathcal I_R(q)[\mathrm{any}\times\mathrm{chiral_{\bar h}}]=\mathrm{Tr}_L(-1)^{2(L_0-\bar h)}q^{L_0},
\end{equation}
where the trace is taken over the left-moving representation.  For $\mathcal N=(2,2)$ supersymmetry, the only non-vanishing contributions are then
\begin{align}
\mathcal{I}_R(q)[\mathrm{long}_{h,r}\times\mathrm{chiral}_{\bar{h}}]&=(-1)^{2(h-\bar{h})}\frac{(1-\sqrt{q})^2
    q^h}{1-q},\nn\\
\mathcal{I}_R(q)[\mathrm{short}_h\times\mathrm{chiral}_{\bar{h}}]&=(-1)^{2(h-\bar{h})}\frac{(1-\sqrt{q})q^h}{1-q},
\end{align}
where the short representation on the left can be either chiral or antichiral.  It is now straightforward to check that the prescription (\ref{eq:mainResult}) acting on $\mathcal I_R(q)$ successfully reproduces the result of Table~\ref{tbl:(2,2)SUSY}.

\subsection{The Small \texorpdfstring{$\mathcal{N}=4$}{N=4} Superconformal Algebra}\label{subsec:N4rep}

The $\mathcal{N}=4$ superconformal algebra corresponds to PSU($1,1|2$), with bosonic
subgroup SU$(1,1)\times$SU($2$). The global part of the algebra is
given by $L_0$, $L_{\pm1}$, $G_{\pm1/2}^a$, $\tilde G_{\pm1/2}^a$
and $J_0^i$.  Lowest weight representations are built from the state
$|h,j,m=j\rangle$, where $h$ is the $L_0$ eigenvalue and $j,m$ are
the $SU(2)$ quantum numbers.  This state is annihilated by $L_1$,
$G_{1/2}^a$, $\tilde G_{1/2}^a$, $J_1^i$ and $J_0^+$.  Descendants
are built using the creation operators $L_{-1}$, $G_{-1/2}^a$,
$\tilde G_{-1/2}^a$ as well as the SU($2$) generators $J_0^i$.

\subsubsection*{Representations of the Small $\mathcal{N}=4$ Superconformal Algebra}

Representations come in two forms, long (massive, with $h>j$) and
short (massless, with $h=j$).  The long representations are generated by acting with all four
supercharge creation operators $G_{-1/2}^a$ and $\tilde G_{-1/2}^a$,
transforming as $\mathbf{2}+\bar{\mathbf{2}}$ of SU($2)$.  A generic
long representation is built from $16$ $\mathcal N=0$ representations,
all transforming as representations of SU$(2)$
\begin{align}
&|h,j\rangle,\nn\\
&2|h+\ft12,j+\ft12\rangle,&&2|h+\ft12,j-\ft12\rangle,\nn\\
&|h+1,j+1\rangle,&&4|h+1,j\rangle,&&|h+1,j-1\rangle,\nn\\
&2|h+\ft32,j+\ft12\rangle,&&2|h+\ft32,j-\ft12\rangle,\nn\\
&|h+2,j\rangle.
\end{align}
These representations decompose into a number of long
$\mathcal{N}=2$ multiplets, and hence do not contribute to the index.

A short representation of the $\mathcal{N}=4$ superconformal algebra has $h=j$.
Starting from $|j,j,m=j\rangle$, we can see that $G_{-1/2}^2$ and
$\tilde G_{-1/2}^1$ generate null states.  We thus use $G_{-1/2}^1$
and $\tilde G_{-1/2}^2$ to generate descendants.  The short
multiplet contains
\begin{equation}
|j,j\rangle,\qquad2|j+\ft12,j-\ft12\rangle,\qquad|j+1,j-1\rangle.
\end{equation}
In particular, these short representations decompose into one chiral,
one antichiral, and a number of long $\mathcal{N}=2$ multiplets.
Therefore if such an $\mathcal N=4$ short representation is present on
the right sector, it contributes to $\mathcal{I}_R$ just as its
constituent $\mathcal N=2$ chiral multiplet does, namely $(-1)^{-2\bar{h}}$.

\subsubsection*{Contribution to the Index}
The right-handed index only receives contributions from shortened representations on the right
\begin{equation}
\mathcal I_R(q)[\mathrm{any}\times\mathrm{short_{\bar h}}]=\mathrm{Tr}_L(-1)^{2(L_0-\bar h)}q^{L_0}.
\end{equation}
For the non-chiral $\mathcal N=(4,4)$ algebra, we have
\begin{align}
\mathcal{I}_R(q)[\mathrm{long}_{h,r}\times\mathrm{short}_{\bar{h}}]=(-1)^{2(h-\bar{h})}\frac{(2h+1)(1-\sqrt{q})^3
    q^h}{1+\sqrt{q}},\nn\\
\mathcal{I}_R(q)[\mathrm{short}_h\times\mathrm{short}_{\bar{h}}]=(-1)^{2(h-\bar{h})}\frac{(1+\sqrt{q}+2h(1-\sqrt{q}))q^h}{1+\sqrt{q}}.
\end{align}
The short$_{h}\times$short$_{\bar{h}}$ representations of the
$\mathcal{N}=(4,4)$ algebra are sometimes denoted as
$(h,\bar{h})_s$.
It is straightforward to check that the prescription
(\ref{eq:mainResult}) acting on the above successfully reproduces the
result of Table~\ref{tbl:(4,4)SUSY}.

\subsection{The Large \texorpdfstring{$\mathcal{N}=4$}{N=4} Superconformal Algebra}

The bulk supergravity symmetry group for the duality involving the
large $\mathcal{N}=(4,4)$ SCA is $D(2,1|\alpha)\times
D(2,1|\alpha)$. Therefore we focus our discussion on a single copy of
$D(2,1|\alpha)$.
The bosonic subgroup of $D(2,1|\alpha)$ is SL($2,R$)$\times$
SU($2$)$\times$SU($2$). Therefore three quantum numbers $h,\ j^+,\ j^-$ will be
used to label its lowest weight representations. Note that the
large $\mathcal{N}=4$ algebra has an extra U($1$), but only the
singlet sector of that U($1$) is of interest to us. Hence the
corresponding quantum number never appears in our discussion.

For comparison with the better known $\mathcal{N}=2$ algebra, note
that $D(2,1|\alpha)$ has an $\mathcal{N}=2$ subgroup, whose
SL$(2,R)$ part coincides with that of $D(2,1|\alpha)$, and whose
U($1$) part is given by the combination
\begin{equation}
J^{\mathcal{N}=2}=\gamma J_z^+ + (1-\gamma)J_z^-,
\label{eq:JofN=2}
\end{equation}
of the third-component generators of the two SU($2$)'s inside $D(2,1|\alpha)$, where $\gamma=\alpha/(1+\alpha)$.

\subsubsection*{Representations of $D(2,1|\alpha)$}
Unitary representations of $D(2,1|\alpha)$ satisfy the BPS inequality
\begin{equation}
h\ge\gamma j^+ + (1-\gamma)j^-,
\end{equation}
with short representations saturating the bound.  The long representations of $D(2,1|\alpha)$ are shown in
Table~\ref{tbl:longDrep}, and the short representations are shown in Table~\ref{tbl:shortDrep}.  Note that only short representations with $j^+=j^-=j$ appear in the spectrum of string theory on AdS$_3\times S^3\times S^3\times S^1$ \cite{Eberhardt:2017a,Baggio:2017kza}. For such representations,  the BPS relation becomes $h=j$.

\begin{table}[t]
    \begin{center}
        \begin{tabular}{llll}
            $|h,j^+,j^-\rangle$,\\
            $|h+\fft12,j^+ + \frac{1}{2},j^- + \frac{1}{2}\rangle$, & $|h+\fft12,j^+ + \frac{1}{2},j^- - \frac{1}{2}\rangle$, & $|h+\fft12,j^+ - \frac{1}{2},j^- + \frac{1}{2}\rangle$, & $|h+\fft12,j^+ - \frac{1}{2},j^- -\frac{1}{2}\rangle$,\\
            \multispan4\kern.5em$|h+1,j^+ +1,j^-\rangle$,  $|h+1,j^+,j^- +1\rangle$, $2|h+1,j^+,j^-\rangle$,  $|h+1,j^+ -1,j^-\rangle$, $|h+1,j^+,j^- -1\rangle$,\\
            $|h+\fft32,j^+ + \frac{1}{2},j^- + \frac{1}{2}\rangle$, & $|h+\fft32,j^+ + \frac{1}{2},j^- - \frac{1}{2}\rangle$, & $|h+\fft32j^+ - \frac{1}{2},j^- + \frac{1}{2}\rangle$, & $|h+\fft32,j^+ - \frac{1}{2},j^- -\frac{1}{2}\rangle$,\\
            $|h+2,j^+,j^-\rangle$.
        \end{tabular}
        \caption{The long representations of $D(2,1|\alpha)$, where $h>\gamma j^+ + (1-\gamma)j^-$.}\label{tbl:longDrep}
    \end{center}
\end{table}

\begin{table}[t]
    \begin{center}
        \begin{tabular}{lll}
            $|h,j^+,j^-\rangle$,\\
            $|h+\fft12,j^+ + \frac{1}{2},j^- - \frac{1}{2}\rangle$, & $|h+\fft12,j^+ - \frac{1}{2},j^- + \frac{1}{2}\rangle$, & $|h+\fft12,j^+ - \frac{1}{2},j^- -\frac{1}{2}\rangle$,\\
            $|h+1,j^+,j^-\rangle$, & $|h+1,j^+ -1,j^-\rangle$, & $|h+1,j^+,j^- -1\rangle$,\\
            $|h+\fft32,j^+ - \frac{1}{2},j^- -\frac{1}{2}\rangle$.
        \end{tabular}
        \caption{The short representations of $D(2,1|\alpha)$, where $h=\gamma j^+ + (1-\gamma)j^-$.}\label{tbl:shortDrep}
    \end{center}
\end{table}

\subsubsection*{Contribution to the Index}

Long representations of $D(2,1|\alpha)$ can be decomposed into a number of long $\mathcal N=2$ multiplets, and hence do not contribute to the right-handed superconformal index.  On the other hand, short representations of $D(2,1|\alpha)$ can yield short $\mathcal N=2$ multiplets, and hence can contribute to the index.  To see how this arises, we compare the $D(2,1|\alpha)$ shortening condition, $h=\gamma j^+ + (1-\gamma)j^-$, with the $\mathcal N=2$ shortening condition, $h=|r|$.  Using the relation (\ref{eq:JofN=2}) between the $\mathcal N=2$ current and the $D(2,1|\alpha)$ currents then demonstrates that a short $D(2,1|\alpha)$ representation contains one chiral and one anti-chiral $\mathcal N=2$ multiplet, along with a set of long $\mathcal N=2$ multiplets.  The chiral (anti-chiral) multiplet corresponds to the state within $|h,j^+,j^-\rangle$ where the third components of the two SU($2$)'s are both maximal (minimal).

As a result, the contribution of a short $D(2,1|\alpha)$ multiplet on the right to $\mathcal I_R(q)$ takes the form
\begin{equation}
\mathcal I_R(q)[\mathrm{any}\times\mathrm{short_{\bar j^+,\bar j^-}}]=\mathrm{Tr}_L(-1)^{2(L_0-\bar h)}q^{L_0}.
\end{equation}
The contribution of $D(2,1|\alpha)\times D(2,1|\alpha)$ representations to the right-handed index can then be determined to be
\begin{align}
\mathcal{I}_R(q)[\mathrm{long}_{h,j^+,j^-}\times\mathrm{short}_{\bar{j}^+,\bar{j}^-}]&=\frac{(-1)^{2(h-\bar{h})}(2j^+
    +1)(2j^- +1)(1-\sqrt{q})^4 q^h}{1-q},\nn\\
\mathcal{I}_R(q)[\mathrm{short}_{j^+,j^-}\times \mathrm{short}_{\bar{j}^+,\bar{j}^-}]&\nn\\
&\kern-4em=\frac{(-1)^{2(h-\bar{h})}(1+2(j^+ +j^-)(1-\sqrt{q})+4j^+ j^-
    (1-\sqrt{q})^2+\sqrt{q}) q^h}{1+\sqrt{q}}.
\end{align}
Once again, we can verify that (\ref{eq:mainResult}) acting the index successfully reproduces the result of
Table~\ref{tbl:large(4,4)SUSY}.

\section{Zeta Function Representation of the \texorpdfstring{Large-$c$}{Large-c} Index and the Supersymmetric Casimir Energy}\label{App:zeta}

In the main text we have focused on the relation between the
supersymmetric Casimir energy and the single-particle supersymmetric
index. The single-particle index is advantageous for practical
computations but the full multi-particle index would perhaps be conceptually more
satisfying. Happily, the relation between the two is particularly
simple in the large-$c$ limit. In this appendix we show how the
supersymmetric Casimir energy as given in (\ref{eq:sumExpressionEL})
naturally appears as a linear-in-$\beta$ term in the natural
logarithm of the full multi-particle large-$c$ index. Additionally, we show how this term is related to the standard definition of the Casimir
energy in terms of a regulated sum over energy eigenvalues which
appears as the zero temperature contribution to the free energy. We stress that we are not
presenting a CFT$_2$ discussion in this appendix, but studying the
AdS$_3$ side.

\subsection{Supersymmetric Casimir Energy from the Index}

Starting from the large-$c$ single-particle index in \eqref{eq:IstR}, the full index is given by enumerating all multi-particle states which arise as products of single-particle states. This is accomplished by taking the Plethystic exponential as in \eqref{eq:IRPE}, which we reproduce here
\begin{equation}\label{eq:largecI-app}
\mathcal I^\text{large-c}_R(q) = \exp \left( \sum\limits_{k=1}^\infty\frac{1}{k} \mathcal I_R^{s.p.}(q^k)\right).
\end{equation}
To simplify the notation we will omit the large-c superscript in the following.

For concreteness, consider an $\mathcal N = (0,2)$ chiral multiplet (cases with higher amounts of supersymmetry can be treated similarly).
The single-particle index \eqref{eq:I(0,2)} of such a chiral multiplet is
\begin{align}
\mathcal I^{s.p.}_{R,h,\bar h}(q) &= (-1)^{2(h-\bar h)}\frac{q^h}{1-q}= (-1)^{2(h-\bar h)} \sum_{n=0}^\infty q^{h+n}~.
\end{align}
Inserting this into \eqref{eq:largecI-app}, the natural logarithm of the large-$c$ index is given by the series
\begin{equation}\label{eq:IRhh-app}
\ln \mathcal I_{R,h,\bar h}(e^{-\beta}) = (-1)^{2(h-\bar h)}\sum_{k=1}^\infty \sum_{n=0}^\infty \frac{1}{k}  e^{-k\beta(h+n)},
\end{equation}
where we have set $q=e^{-\beta}.$

In order to extract the
small-$\beta$ asymptotics of this expression, we follow Cardy
\cite{Cardy:1991kr} and write the exponential as the inverse Mellin
transform of the Gamma function
\begin{equation}
e^{-x} = \frac{1}{2\pi i} \int_{\mathcal C} x^{-s}\Gamma(s)\, ds,
\end{equation}
where the contour runs to the right of and parallel to the imaginary axis. Inserting this into the logarithm of the index we have
\begin{equation}\label{eq:lnIMT-app}
\ln \mathcal I_{R,h,\bar h}(e^{-\beta})  = (-1)^{2(h-\bar h)}\frac{1}{2\pi i}\int_{\mathcal C} \beta^{-s} \zeta(s+1) \Gamma(s)\sum_{n=0}^\infty (h+n)^{-s}\, ds,
\end{equation}
where $\zeta(s+1)$ comes from the sum over $k$ in
\eqref{eq:IRhh-app}. The contour is now to the right of
$\mathrm{Re}s=1$ where the integrand is convergent. This contour can be closed by deforming the upper ($\mathrm{Im}\,s >0$) and lower ($\mathrm{Im}\,s < 0$) segments up and down towards the negative real axis from above and below. Cauchy's theorem then gives the small-$\beta$ expansion
\begin{equation}\label{eq:Esusyzeta}
\ln \mathcal I_{R,h,\bar h}(e^{-\beta}) = (-1)^{2(h-\bar h)}\left( \frac{1}{\beta} \frac{\pi^2}{6} +   \zeta'(h;0) +
\beta\frac{1}{2} \zeta(h;-1)+\cdots \right),
\end{equation}
where
\begin{equation}
\zeta(h;s) = \sum_{n=0}^\infty (h+n)^{-s},
\end{equation}
is a Hurwitz zeta function. The dots in \eqref{eq:Esusyzeta} refer
to higher powers in $\beta$ that arise from the infinite set of
poles in the $\Gamma$-function at $s=-n$ for integer $n>1$. These
terms are subleading for small $\beta.$

We are particularly interested in the term that is linear in
$\beta$. It is due to the pole of the integrand in
\eqref{eq:lnIMT-app} at $s=-1$ which arises from the
$\Gamma$-function. To find its coefficient we have used
$\mathrm{Res}(\Gamma,-1) = -1$ as well as
$\zeta(0)=-1/2.$\footnote{To get the other terms shown in
\eqref{eq:Esusyzeta} we have used $\mathrm{Res}(\Gamma,0) = 1$ and
$\zeta(2) = \pi^2/6$ for the $\beta^{-1}$ term. For the $\beta^0$
term, we note that the Laurent expansion of $\zeta(s+1)\Gamma(s) =
1/s^2 + \mathcal{O}(s^0)$ near $s=0,$ which gives rise to the
derivative $\zeta'(0;h) \equiv \frac{\partial}{\partial s}
\zeta(s;h)|_{s=0}$ appearing in the residue of the pole at this
point.}
We now recognize that the coefficient of $\beta$ in
\eqref{eq:Esusyzeta} is nothing but the sum
(\ref{eq:sumExpressionEL}) over the left-handed quantum
zero-energies of the supersymmetric AdS$_3$ modes in the multiplet,
and thus it gives $E_{(1)}^L(h,\bar h)$ as claimed. Of course, the
Hurwitz zeta function that appears in this term is elementary so we
can evaluate the coefficient of $\beta$ in \eqref{eq:Esusyzeta}
explicitly and find
\begin{equation}
\label{eq:EL1zeta} \frac{1}{2}(-1)^{2(h-\bar h)} \zeta(h ; -1) =
-\frac{1}{24}(-1)^{2(h-\bar h)} (1 - 6 h +6 h^2)~.
\end{equation}
This matches our result in \eqref{eq:chiral_Esusy} that was derived
from the single-particle index using the prescription
\eqref{eq:EL1index}.

That the coefficient of $\beta$ in the small-$\beta$ expansion of
$\ln \mathcal{I}_R$ should be $E_{(1)}^L$ follows from our
prescription (\ref{eq:EL1index}) as well; the reader might convince
themselves of this using (\ref{eq:largecI-app}), or more carefully
using the relation between the small-$\beta$ asymptotics of the
single-particle and multi-particle indices presented in Appendix~D
of \cite{Ardehali:thesis}.\\


To summarize this appendix so far, we have shown that
\begin{itemize}
\item
the supersymmetric Casimir energy $E_{(1)}^L$ that we obtained from
a single-particle-index regularization (\ref{eq:EL1index}) of the
expression (\ref{eq:sumExpressionEL}), also arises as the
coefficient of $\beta$ in the small-$\beta$ asymptotic expansion of
logarithm of the bulk multi-particle index;
\item
at the level of individual bulk multiplets, the index prescription
(\ref{eq:EL1index}) and the zeta-function regularization
(\ref{eq:EL1zeta}) amount to the same result for the expression
(\ref{eq:sumExpressionEL}).
\end{itemize}

%

\subsection{Supersymmetric Casimir Energy as a Sum over Normal Mode Energies}

We now want to show that the SUSY Casimir energy in
\eqref{eq:EL1zeta} corresponds to a sum over \emph{supersymmetric
energies} of states in the bulk chiral multiplet.

To see this note that a field  in global AdS$_3$ with conformal weight $(h, \bar h)$ has quantized normal mode energies \cite{Zhang:2012kya}
\begin{equation}\label{eq:normalmodes}
E_{n,l}(h,\bar h) = \frac{2n+ |l| + h+\bar h}{\ell_3}, \qquad n = 0,\,1,\,2,\,\dots \,, \,\,\, l = 0,\,\pm1,\,\pm2,\,\dots\,,
\end{equation}
where $l$ is the ``angular" momentum around the spatial circle, $n$ is a radial mode number and $\ell_3$ is the AdS radius. We will set $\ell_3=1$
in the following.

In the supersymmetric context that we consider, these energies are
modified due to the presence of a constant background $U(1)_R$ gauge
field that is required in order for the bulk partition function in
global AdS$_3$ to correspond to the supersymmetric partition
function (or elliptic genus) \cite{Kraus:2006p}.\footnote{See also
    \cite{Genolini:2016ecx,Genolini:2016sxe} for the corresponding
    statements in AdS$_5$.} This constant background
$U(1)_R$ gauge field is a flat connection so it simply shifts the energies by an amount equal
to the $R$-charge such that the supersymmetric energies become
\begin{equation}
E^{SUSY}_{n,l}(h,\bar h;j) = 2n+ |l| + h+\bar h - \bar
r,\label{eq:Enl}
\end{equation}
where $n$ and $l$ have the same ranges as in \eqref{eq:normalmodes} and $\bar r$ is the $U(1)_R$ charge of the field.

Each entry of the oscillator spectrum (\ref{eq:Enl}) contributes the
usual vacuum energy ${1\over 2}\hbar\omega$. The supersymmetric
Casimir energy follows as a renormalized sum over these
contributions so using zeta-function regularization we find:
\begin{equation}\label{eq:Esusysum}
E^{SUSY}(h,\bar h;\bar{r}) := \frac{1}{2}(-1)^{2(h-\bar
h)}\sum\limits_{\kappa=0}^\infty (\kappa+1)(\kappa+ h+\bar h - \bar
r)^{-s}\Big|_{s=-1}~.
\end{equation}
We have inserted an extra $(-1)^{2(h-\bar h)}$ in order to
count bosons with a plus sign and fermions with a minus sign. Also,
in \eqref{eq:Esusysum} we have replaced the double sum on $n$ and
$l$ with a single sum that takes into account the appropriate
degeneracy of each energy level.

For the $\mathcal N=(0,2)$ chiral multiplet in \eqref{eq:N=02ch},
the $U(1)_R$ charge is related to the weight by $\bar r=\bar h$, so
we have
\begin{equation}
E^{ch}(h,\bar h;\bar{r}=\bar h) = \frac{1}{2}(-1)^{2(h-\bar
h)}\sum\limits_{\kappa=0}^\infty\left[ (\kappa+1)(\kappa+ h)^{-s} -
(\kappa+1)(\kappa+ h + 1)^{-s} \right]_{s=-1}.
\end{equation}
After shifting the summation index in the second term, this can be written
\begin{equation}\label{eq:Esusych}
E^{ch}(h,\bar h) = \frac{1}{2}(-1)^{2(h-\bar h)}\sum\limits_{\kappa=0}^\infty (\kappa+ h)^{-s} \Big|_{s=-1},
\end{equation}
which is equal to the SUSY Casimir energy defined from the index in
\eqref{eq:Esusyzeta}.

On the other hand, one can check that for the long $\mathcal N =
(0,2)$ multiplet in \eqref{eq:N=02long} the sum over the
supersymmetric energies vanishes due to cancellation between bosons
and fermions.


\section{Computation of One-loop Chern-Simons Terms} \label{App:CS}

In this appendix we collect some details of the computation of the
Chern-Simons levels claimed in \eqref{eq:CSLag} of
Section~\ref{sec:Index}, for spins less than two. It amounts to
computing gauge boson vacuum polarization diagrams in flat space
with charged fields running in the loop and extracting the
parity-odd term that survives in the low-energy limit. We proceed
directly to the vacuum polarization computation and only then
providing the full context. The main result is given in equation
\eqref{eq:keff}, where the spin $s$ is given from the results below
by identifying $\text{sign}(m)  = \frac{s}{|s|}$
\cite{Tyutin:1997yn}.

The generation of Chern-Simons couplings in the three-dimensional effective action from one-loop vacuum polarization diagrams due to spin-$1/2$ fermions was first found in \cite{Redlich:1983kn,Redlich:1983dv}, which we generalize to fields of spin-$1$ and spin-$3/2$ and conjecture an extrapolation to higher spins. The analysis of this section follows the methods and regularization of \cite{Bonetti:2013ela} where the analogous calculation in five dimensions is performed.

\subsection{Details on the Vacuum Polarization Diagrams}

The integrand of the one-loop diagrams we will consider are all of the form
\begin{equation}
\Sigma^{\mu\nu}(p) = \int \frac{d^3k}{(2\pi)^3} \frac{\mathbb N^{\mu\nu}(p,k;m)}{(k^2+m^2)((k-p)^2+m^2)},
\end{equation}
where $p^\mu$ is the external photon momentum and $k^\mu$ is the loop momentum.

Here we will go through some of the details in extracting the relevant terms in the numerator factors. In particular, since we are interested in extracting the parity odd term we can restrict to the terms in the numerator of the form
\begin{equation}
\mathbb N^{\mu\nu}(p,k;m) = f(k^2;m) \epsilon^{\mu\nu\rho}p_\rho.
\end{equation}
In particular, we can discard terms with higher powers of $p$ as well as those that do not include the Levi-Civita tensor.

\subsubsection{\texorpdfstring{Spin-$1/2$}{spin-1/2}}

For fermions, the relevant parity odd terms arise from traces of $\gamma$ matrices.%

The spin-$1/2$ propagator is
\begin{equation}
\Delta^{1/2}(p) = \frac{-\slashed{p}+im\,\,}{p^2+m^2}
\end{equation}
and the interaction vertex is simply
\begin{equation}
V_{1/2}^{\mu} = q \gamma^\mu.
\end{equation}
From these, we construct the numerator for a spin-$1/2$ field in the loop to be
\begin{eqnarray}
\mathbb N^{\mu\nu}_{1/2} (p,k;m) &=& - \text{Tr}\left[(-q\gamma^\mu) (-\slashed{k}+i m) (-q\gamma^\nu)(-(\slashed{k}-\slashed{p})+m)\right] \nn \\
&\simeq& q^2 \left[ -\text{Tr}(\gamma^\mu\gamma^\rho\gamma^\nu)(im k_\rho) -\text{Tr}(\gamma^\mu\gamma^\nu\gamma^\rho)(im (k_\rho - p_\rho))\right] \nn \\
&\simeq& 2imq^2 \left[ -\epsilon^{\mu\rho\nu} k_\rho -\epsilon^{\mu\nu\rho}(k_\rho - p_\rho)\right] \nn\\
&\simeq& 2imq^2 \epsilon^{\mu\nu\rho} p_\rho.
\end{eqnarray}
Putting this all together we find the following parity-odd contribution to the effective action
\begin{eqnarray}
\Sigma_{1/2}^{\mu\nu} e_{\mu}e_{\nu}&= & (-) \int \frac{d^3k}{(2\pi)^3} \text{Tr} \left( (-q \gamma^\mu) \frac{-\slashed{k}+im}{k^2+m^2} (-q\gamma^\nu)\frac{-(\slashed{k}-\slashed{p})+im}{(k-p)^2+m^2} \right) \nn\\
&\simeq& -2i m q^2 p_{\rho}e_\mu e_\nu \epsilon^{\rho\mu\nu} \int \frac{d^3k}{(2\pi)^3}\frac{1}{(k^2+m^2)((k-p)^2+m^2)} \nn \\
&\simeq& -\frac{i}{2\pi^2}  \frac{m}{|m|}q^2\Gamma\left(\ft32\right)\Gamma\left(\ft12,\Lambda^{-2}\right) p_{\rho}e_\mu e_\nu \epsilon^{\rho\mu\nu} \nn \\
&\simeq& \frac{i}{4\pi}  \text{sign}(m) q^2 p_{\rho}e_\mu e_\nu \epsilon^{\rho\mu\nu}\,.
\end{eqnarray}

\subsubsection{\texorpdfstring{Spin-$1$}{spin-1}}

For bosons, the parity odd term arises from explicit $\epsilon$ tensors in both the propagator and the vertex. For spin-$1,$ these are
\begin{eqnarray}
\Delta^1_{\mu\nu}(p) &=& -i\left(m \eta_{\mu\nu}+\ft{1}{m}p_\mu p_\nu - i \epsilon_{\mu\nu\gamma}p^\gamma\right), \nn\\
V_1^{\nu\mu\lambda} &=&  - q \epsilon^{\nu\mu\lambda}.
\end{eqnarray}
These give the numerator
\begin{eqnarray}\label{eq:spin1N}
\mathbb N^{\mu\nu}_{1} (p,k;m) &=& (- q \epsilon^{\lambda\mu\sigma})\Delta^1_{\lambda\rho}(k) (- q \epsilon^{\rho\nu\gamma}) \Delta^1_{\gamma\sigma}(k-p),\nn \\
&\simeq& -i m q^2 \epsilon^{\mu\nu\sigma}p_{\sigma}\left(1 - \frac{k^2}{m^2}\right).
\end{eqnarray}
From which we compute
\begin{eqnarray}
\Sigma_{1}^{\mu\nu} e_{\mu}e_{\nu} &=& (+) \int \frac{d^3k}{(2\pi)^3} \frac{ (- q \epsilon^{\lambda\mu\sigma})\Delta^1_{\lambda\rho}(k) (- q \epsilon^{\rho\nu\gamma}) \Delta^1_{\gamma\sigma}(k-p)}{(k^2+m^2)((k-p)^2+m^2)}\nn\\
&\simeq& -i m q^2 p_{\rho}e_\mu e_\nu \epsilon^{\rho\mu\nu} \int \frac{d^3k}{(2\pi)^3}\left(1-\frac{k^2}{m^2}\right)\frac{1}{(k^2+m^2)((k-p)^2+m^2)} \nn\\
&\simeq& -\frac{i}{4\pi^2}  \frac{m}{|m|}q^2\bigg(\Gamma\left(\ft32\right)\Gamma\left(\ft12,\Lambda^{-2}\right)-\Gamma\left(\ft52\right)\Gamma\left(-\ft12,\Lambda^{-2}\right)\bigg) p_{\rho}e_\mu e_\nu \epsilon^{\rho\mu\nu} \nn \\
&\simeq& \frac{i}{2\pi}\left(\frac{3\Lambda}{4\sqrt{\pi}} - 1\right)\text{sign}(m)q^2 p_{\rho}e_\mu e_\nu \epsilon^{\rho\mu\nu}\,.
\end{eqnarray}

\subsubsection{\texorpdfstring{Spin-$3/2$}{spin-3/2}}

For massive spin-$3/2$ fields in the loop, we have
\begin{eqnarray}\label{eq:spin32prop}
\Delta^{3/2}_{\mu\nu}(p) &=& \left(\eta_{\mu\nu} + \frac{p_\mu p_\nu}{m^2}\right)(-\slashed{p} + im) + \frac{1}{2}\left(\gamma_\mu - \frac{ip_\mu}{m}\right)\left(-\slashed{p}-im\right)\left(\gamma_\nu - \frac{ip_\nu}{m}\right)\,,\nn\\
V^{\nu\mu\lambda} &=& - q \gamma^{\nu\mu\lambda}.
\end{eqnarray}
These give a numerator of
\begin{eqnarray}
\mathbb N^{\mu\nu}_{3/2} (p,k;m) &=& -\text{Tr}\left[(- q \epsilon^{\lambda\mu\sigma})\Delta^{\frac32}_{\lambda\rho}(k) (- q \epsilon^{\rho\nu\gamma}) \Delta^{\frac32}_{\gamma\sigma}(k-p)\right].
\end{eqnarray}
To simplify this, schematically write
\begin{equation}
\Delta^{\frac32}_{\mu\nu}(p) = A_{\mu\nu}(p) + B_{\mu\nu}(p),
\end{equation}
where $A_{\mu\nu}(p)$ refers to the first term in (\ref{eq:spin32prop}) and $B_{\mu\nu}(p)$ to the second. The numerator can then be written
\begin{eqnarray}
\mathbb N^{\mu\nu}_{3/2} (p,k;m) &=& - q^2\epsilon^{\lambda\mu\sigma}\epsilon^{\rho\nu\gamma}\text{Tr}\left[ \big(A_{\lambda\rho}(k) + B_{\lambda\rho}(k)\big)\big(A_{\gamma\sigma}(k-p) + B_{\gamma\sigma}(k-p)\big)\right].
\end{eqnarray}
Evaluating the various terms in the product independently we find
\begin{eqnarray}
\epsilon^{\lambda\mu\sigma}\epsilon^{\rho\nu\gamma}\text{Tr}\left[A_{\lambda\rho}(k)A_{\gamma\sigma}(k-p)\right] &\simeq& 0, \nn \\
\epsilon^{\lambda\mu\sigma}\epsilon^{\rho\nu\gamma}\text{Tr}\left[A_{\lambda\rho}(k)B_{\gamma\sigma}(k-p)\right] &\simeq& -i m \epsilon^{\mu\nu\rho} p_\rho \left(1 + \frac{k^2}{m^2}\right), \nn\\ \epsilon^{\lambda\mu\sigma}\epsilon^{\rho\nu\gamma}\text{Tr}\left[B_{\lambda\rho}(k)A_{\gamma\sigma}(k-p)\right] &\simeq& -i m \epsilon^{\mu\nu\rho} p_\rho \left(1 + 3 \frac{k^2}{m^2}\right), \nn\\ \epsilon^{\lambda\mu\sigma}\epsilon^{\rho\nu\gamma}\text{Tr}\left[B_{\lambda\rho}(k)B_{\gamma\sigma}(k-p)\right] &\simeq& 2i m \epsilon^{\mu\nu\rho} p_\rho \left(1 + \frac{k^2}{m^2}\right).
\end{eqnarray}
Adding everything together we find the numerator is given by
\begin{eqnarray}
\mathbb N^{\mu\nu}_{3/2} (p,k;m) &=& -2 i m q^2 k^2 \epsilon^{\mu\nu\rho} p_\rho.
\end{eqnarray}
Putting it altogether, we find the effective parity odd term
\begin{eqnarray}
\Sigma^{\mu\nu}_{3/2}e_{\mu}e_{\nu}&= & (-) \int \frac{d^3k}{(2\pi)^3} \frac{\text{Tr}\left[(-q\epsilon^{\delta\mu\rho}) \Delta^{\frac{3}{2}}_{\delta\sigma}(k) (-q\epsilon^{\lambda\nu\sigma})\Delta^{\frac{3}{2}}_{\lambda\rho}(k-p)\right]}{(k^2+m^2)((k-p)^2+m^2)} \nn\\
&\simeq& -2 i m q^2 p_{\rho}e_\mu e_\nu \epsilon^{\rho\mu\nu} \int \frac{d^3k}{(2\pi)^3}\frac{k^2}{m^2}\frac{1}{(k^2+m^2)((k-p)^2+m^2)} \nn \\
&\simeq& -\frac{i}{2\pi^2} \frac{m}{|m|}q^2\Gamma\left(\ft52\right)\Gamma\left(-\ft12,\Lambda^{-2}\right)p_{\rho}e_\mu e_\nu \epsilon^{\rho\mu\nu} \nn \\
&\simeq& -\frac{i}{2\pi}\left(\frac{3\Lambda}{2\sqrt{\pi}} - \frac{3}{2}\right)\text{sign}(m)q^2p_{\rho}e_\mu e_\nu \epsilon^{\rho\mu\nu}\,.
\end{eqnarray}

\subsubsection{Extrapolation to Arbitrary Spin}

The results above all have a finite term given by
\begin{equation}
\Sigma^{\mu\nu}_{|s|} e_\mu e_\nu = -(-1)^{2s}\frac{i}{2\pi} s \,q^2 p_\rho e_\mu e_\nu \epsilon^{\rho\mu\nu},
\end{equation}
where $s$ is the spin of the field appearing in the loop, which is
related to the mass by $m = \frac{s}{|s|}.$ Given the chiral nature
of the massive higher spin fields in three-dimensions
\cite{Tyutin:1997yn}, it is reasonable to suspect that charged
massive higher spin fields will continue to follow this pattern.
This suspicion is corroborated by the results in
Section~\ref{sec:Index}. It would be very interesting to verify this
conjectured structure with an explicit calculation, but we will
leave this for future work.

\subsection{Momentum Integrals}

In order to compute the momentum integrals we have utilized the Schwinger parametrization of the integrand. Here are some expressions relevant for this.

Since we are only interested in terms in the effective action of the form $p_{\rho}e_\mu e_\nu \epsilon^{\rho\mu\nu},$ we can restrict to the terms from the loop integrals which are independent of the external momentum $p.$

One can rewrite the denominator as
\begin{equation}
\frac{1}{(k^2+m^2)((k-p)^2+m^2)} = \int_0^\infty du \int_0^\infty dv \,\, e^{-(u+v)(\ell^2+\bar m^2)},
\end{equation}
where we have defined the shifted momentum $\ell$ and the shifted mass $\bar m^2$ as
\begin{eqnarray}
\ell &=& k - \frac{v}{u+v}p, \nn \\
\bar m^2&=& m^2 + \frac{uv}{(u+v)^2}p^2.
\end{eqnarray}
Since we are interested in the $p$-independent part of the integral, we drop the $p$ dependence of these.

The integral over the momentum can now be computed by evaluating the Gaussian integral
\begin{eqnarray}
\int \frac{d^3\ell}{(2\pi)^3} \ell^{2n} e^{-\frac{x \ell^2}{m^2}} = \frac{|m|^{2n+3}}{4\pi^2} \frac{\Gamma\left(n+\ft32\right)}{(u+v)^{n+3/2}}.
\end{eqnarray}
Putting this all together we have
\begin{eqnarray}
\int \frac{d^3k}{(2\pi)^3}\frac{k^{2n}}{(k^2+m^2)((k-p)^2+m^2)} &=& \frac{1}{2\pi^2}\frac{m^{2n}}{|m|}\Gamma\left(n+\ft32\right)\int du\,dv \ \Theta\left(u+v - \Lambda^{-2}\right)\frac{e^{-(u+v)}}{(u+v)^{n+3/2}} \nn \\
&=& \frac{1}{4\pi^2}\frac{m^{2n}}{|m|}\Gamma\left(n+\ft32\right)\Gamma\left(-n+ \ft12,\Lambda^{-2}\right).
\end{eqnarray}
In the last line we used
\begin{equation}
\int_0^\infty du\,\int_0^\infty dv\,\,\Theta\left(u+v-\Lambda^{-2}\right)\frac{e^{-(u+v)}}{(u+v)^\alpha} = \Gamma\left(2 - \alpha, \Lambda^{-2}\right).
\end{equation}
The parameter $\Lambda$ is inserted as a UV cut-off. In this way the Schwinger variables act as a heat kernel regulator of the loop integral and $\Lambda$ provides a UV cut-off for integration over the heat variables $u$ and $v.$ One can then perform the integral and expand in large-$\Lambda.$ The regularization procedure is to simply discard the infinite terms proportional to positive powers of $\Lambda.$

\end{document}